\documentclass[epj]{svjour}
\usepackage{graphicx,mathrsfs,amsmath,bm,amssymb}
\usepackage{physics}
\usepackage{hyperref}
\usepackage{multirow}
\usepackage{lscape}
\usepackage{color}
\usepackage{hyperref}
\usepackage{slashed}
\usepackage{aas_macros}
\journalname{Eur. Phys. J. A}

\begin{document}

\title{Primordial black-hole formation and heavy r-process element synthesis from the cosmological QCD transition}
\subtitle{Two aspects of an inhomogeneous early Universe}
\titlerunning{Primordial Black-Hole formation at the Cosmological QCD Transition: Beyond the Standard Model}
\authorrunning{M. Gonin et al.}

\author{
Ma\"el Gonin\inst{1,2}
\and
Günther Hasinger\inst{1,2,3} 
\and
David Blaschke\inst{4,5,6}
\and
Oleksii Ivanytskyi\inst{7} 
\and
Gerd R\"opke\inst{8}
}

\institute{
TU Dresden, Institute of Nuclear and Particle Physics, 01062 Dresden, Germany
\and
Deutsches Zentrum f\"ur Astrophysik (DZA), Postplatz 1, 02826 G\"orlitz, Germany
\and
DESY, Notkestraße 85, 22607 Hamburg, Germany
\and
Helmholtz-Zentrum Dresden-Rossendorf (HZDR), Bautzener Landstrasse 400, 01328 Dresden, Germany
\and
Center for Advanced Systems Understanding (CASUS), Untermarkt 20, 02826 G\"orlitz, Germany
\and
Institute of Theoretical Physics, University of Wroc{\l}aw, 
Max Born place 9, 
50-204 Wroc{\l}aw, Poland 
\and
Centrum for Simulations of Superdense Fluids, University of Wroc{\l}aw, 
Max Born place 9, 50-204 Wroc{\l}aw, Poland 
\and
Institute of Theoretical Physics, University of Rostock, Albert-Einstein-Str. 23, D-18059 Rostock, Germany
}
\date{Received: \today / Revised version: date}
%
\abstract{
We review the role of primordial black holes for illuminating the dark ages of the cosmological evolution and as dark matter candidates.
We elucidate the role of phase transitions for primordial black hole formation in the early Universe and focus our attention to the cosmological QCD phase transition within a recent microscopical model. We explore the impact of physics beyond the Standard Model on the cosmic equation of state and the probability distribution for the formation of primordial black holes which serve as dark-matter candidates. 
We argue that besides primordial black holes also droplet-like quark-gluon plasma inhomogeneities may become gravitationally stabilized for a sufficiently long epoch to distill baryon number and form nuclear matter droplets which upon their evaporation may enrich the cosmos locally with heavy $r$-process elements already in the early Universe.
}
%
%
\maketitle
\section{Introduction}
\label{sec:intro}

Since its first proposal, Big Bang theory, along with the standard model (SM), has been the main pillar of modern cosmology. However, there remain significant question marks in the $\Lambda$ cold dark matter ($\rm \Lambda CDM$) model, such as the nature and origin of Dark Matter (DM). 
As time passed, cosmological, particle, and astrophysical observations have stacked up allowing the scientific community to unravel the Universe, constrain its content, and unveil unexpected conundrums. 
Through these efforts, the DM shadow on the cosmological model starts to light up more and more. As particle DM (pDM) experiments seek for smaller and smaller energy domains, the lack of evidence or even hints of such particle species is stressing the pDM community, pointing towards another possible DM nature. On much bigger scales, through the gravitational wave (GW) observation breakthrough following the detection of the first binary black hole merger event GW150914 in 2015 \cite{LIGOScientific:2016aoc}, and the launch of the James Webb Space Telescope (JWST) in 2021, the astrophysical zoo continued to grow. 
GW astronomy is probably the most striking astrophysical breakthrough of our century, awarded the 2017 Physics Nobel Prize. 
Now we can not only see the Universe content but also hear it chirp. 
The catalog of Black Hole (BH) mergers with now 90 events confirmed by GW observations of the LIGO-Virgo-Kagra (LVK) collaboration \cite{LIGOScientific:2018mvr,LIGOScientific:2020ibl,KAGRA:2021vkt,KAGRA:2021duu} starts the era of statistical studies of BH mergers. 
The JWST can look farther than any other telescope, providing information on galaxies in the early Universe that implies the existence of supermassive black holes (SMBH) and quasars with masses $M\sim10^8 M_\odot$, quite unexpectedly observed at high redshift $z \gtrsim 15$ \cite{Pacucci:2023oci,Goulding:2023gqa,CEERSTeam:2023qgy}.

The idea of linking BH, Big Bang, and DM has been around for more than 50 years since Carr \& Hawking (1974) \cite{Carr:1974nx} introduced the concept of BH being formed right after the big bang, from the collapse of primordial overdensities, such as the so-called primordial black holes (PBH). 
Such a mechanism could form PBH with an extended mass function \cite{Carr:1975qj}, see \cite{Carr:2020xqk,Carr:2020gox,LISACosmologyWorkingGroup:2023njw,Khlopov:2024nqp} for recent reviews and \cite{2024arXiv240605736C} for a historical overview. 
In this context, the JWST and LVK era of observations revitalized the idea of PBH-DM. 
Although PBH could be seeding the SMBH observed by JWST, it has also been argued that at least a fraction of the GW events, including a BH member, could be of primordial origin \cite{Bird:2016dcv,Sasaki:2016jop,Clesse:2016vqa}. 

PBH-DM is a unique probe for new physics, as it combines astronomical observations and very early SM interactions in a unique way, forming BH in the early cosmic quark-gluon plasma (QGP) phase transition. 
The improvement in our understanding of strongly interacting matter through heavy-ion collision experiments \cite{Sorensen:2023zkk}
and lattice QCD simulations \cite{Borsanyi:2016ksw} allows us to draw the cosmic equation of state (EoS) with unprecedented precision over a large range of temperatures, unraveling the universe's thermodynamic history including many cosmic phase transitions. 
Denoting with $p$ the pressure and with $\varepsilon$ the energy density,
the EoS parameter,  
\begin{equation}
\label{eq:EoS}
    w=\frac{p}{\varepsilon},
\end{equation}
and the associated squared speed of sound,
\begin{equation}
\label{eq:css}
    c_s^2=\frac{\partial p}{\partial \varepsilon},
\end{equation}
exhibit a dip at every phase transition of the very early Universe. 
The biggest drop being that of the QCD transition from the QGP phase to the hadron resonance gas \cite{Borsanyi:2016ksw}. 
The dips in $w$ and $c_s^2$ represent a softening of the EoS, which induces an increased probability of gravitational collapse for the largest overdensities, thereby imprinting distinct features in the mass distribution of PBHs \cite{Byrnes:2018clq,Carr:2019kxo,Carr:2020gox,Bodeker:2020stj,Escriva:2022bwe,Musco:2023dak}.

In the present work, we recalculate the Universe's thermal history at the QCD transition using a microscopical model \cite{Blaschke:2023pqd} explained in Sec. \ref{sec:eos}. 
We add for the first time the contribution of the putative X17 particle \cite{Krasznahorkay:2015iga,Krasznahorkay:2021joi,Krasznahorkay:2022pxs,2025arXiv250411439A,Krasznahorky:2024adr,Alves:2023ree} and compare the results with the SM estimates. 
We underline the relevance of such a study in light of many SM anomalies \cite{Crivellin:2025txc}, which make physics beyond the standard model (BSM)  an active field of research with many cosmological consequences and open questions \cite{Khlopov:2024uoo}. 
In Sect. \ref{sec:pbh} we transform the cosmic EoS to a PBH mass distribution.
In Sect. \ref{sec:dm}, we compile recent constraints and circumstantial evidence for PBH-DM, with a focus on the mass range of intermediate-mass black holes (IMBH).
In Sect. \ref{sec:elements}, we consider the 'failed collapse' case following Witten's scenario of a cosmic separation of phases \cite{Witten:1984rs} where baryon number distillation is described within an inhomogenous Big Bang (IBB) and study the implications for the primordial nucleosynthesis of heavy $r$-process elements.
Although these elements are observed in various astrophysical objects, the site where they are formed has not been fully resolved yet.
For an overview on deciphering the origins of the elements through galactic archeology, we refer to the recent review \cite{2025arXiv250318233F} in this EPJA Topical Collection on "Nuclear Physics in Astrophysics", where further references can be found.
Recently, a sudden freeze-out scenario has been considered for the formation of heavy $r$-process elements \cite{2024arXiv241100535R}.
Such a scenario of IBB nucleosynthesis can serve to answer various problems in explaining the abundance of heavy $r$-process nuclei in the earliest objects of the Universe.
In Sect. \ref{sec:summary}, we summarize our findings and present our conclusions.

\section{Cosmic equation of state with QCD transition 
}
\label{sec:eos}
The very early Universe is assumed to be an expanding, extremely hot plasma of elementary particles in thermal equilibrium with radiation. 
As the Universe expands, the temperature drops, and whenever the Universe reaches a (pseudo-) critical temperature of a phase transition, it undergoes a change in its composition. 
We focus our interest on the QCD transition from the QGP to the hadronic matter phase, for which in lattice QCD simulations the pseudo-critical temperature $T_c=156.5$ MeV has been found \cite{HotQCD:2018pds}. 
These ab initio results for the thermodynamics of this transition are equivalently described within an effective microscopical model \cite{Blaschke:2023pqd}.
Within such a unified approach, we evaluate the EoS in the extended temperature range $T\in [1;1300]$ MeV. 
The model uses an ansatz for the thermodynamic potential of QCD, 
\begin{equation} 
\label{eq:omega_total}
    \Omega(T,\mu,\phi,\bar{\phi})=\Omega_{\rm QGP}(T,\mu,\phi,\bar{\phi})+\Omega_{\rm MHRG}(T,\mu,\phi,\bar{\phi}),
\end{equation}
which decomposes it into a QGP sector of quark- and gluon quasiparticles and that of the Mott hadron resonance gas (MHRG). 
In the MHRG model, hadrons are understood as quark bound states (multiquark clusters) that can undergo a Mott dissociation.
We develop that topic in further detail in the Appendix \ref{app:QGP}. 
The following subsection \ref{ssec:CosmicEoS} is about the translation of such a QCD EoS into a cosmic EoS.

\subsection{Cosmic equation of state}
\label{ssec:CosmicEoS} 

Let us start by providing a brief thermal history of the Universe, considering that it starts at the Planck time $t_{\rm Planck}=10^{-43}$ s with a temperature $T_{\rm Planck}=10^{19}$ GeV, when all four fundamental interactions are unified, the Universe is minuscule, and quantum fluctuations are non-negligible. After $t_{\rm Planck}$, gravity decouples and the Grand Unification Era begins. It lasts until strong interactions decouple at $t_{\rm strong}=10^{-36}$ s with $T_{\rm strong} = 10^{16}$ GeV which triggers the inflation era. 
According to Guth's inflationary scenario \cite{Guth:1980zm}, if the Universe undergoes a first-order phase transition with sufficient supercooling during this early epoch, it can enter a period of exponential expansion driven by the energy density of a false vacuum state. 
This inflationary phase addresses two fundamental problems of the standard Big Bang model: the horizon problem and the flatness problem. 
The exponential expansion during inflation stretches quantum fluctuations to cosmological scales, providing seeds for large-scale structure, and possible collapse to PBH.

In the $\Lambda$CDM Standard Model of cosmology, a thermal equilibrium is assumed after the inflation era, when the Universe continues expanding (much slower than during inflation), and reheating starts. The Universe enters the QGP phase and is filled with relativistic SM particles, making the cosmic fluid dominated by radiation.
As expansion continues, the temperature keeps dropping, and the Universe reaches the temperatures $T_i \sim m_i$, when the annihilation of antiparticles with mass $m_i$ is favored over thermal particle-antiparticle creation. Thus, at $T_i$ the antiparticles of species $i$ disappear from the cosmic soup with the consequence that the number of relativistic degrees of freedom $g_*$ drops and the EoS \eqref{eq:EoS} deviates from the value of $w=1/3$ that characterizes a radiation era. 
Hereafter, we call these types of events cosmic phase transitions. We focus our interest on the QCD transition, when the Universe goes from the QGP to the hadron resonance gas phase (see the Appendix \ref{app:QGP}), and quarks and gluons become confined into hadrons and mesons. This transition is of particular interest because it is the moment when the baryon asymmetry freezes out.

The key parameters to determine the early Universe's EoS are: entropy, energy density, and pressure.
These thermodynamic quantities are related as follows.

\begin{equation}\label{eq:PerfectFluid}
s(T)=\frac{dp}{dT}  \quad , \quad  \varepsilon(T)=s(T)T - p(T)\,.
\end{equation}

Various studies explore the scenario with non-vanishing baryon and lepton chemical potentials, \cite{Borsanyi:2021sxv,Middeldorf-Wygas:2020glx,Wygas:2018otj}. Analog quantities like the number of relativistic degrees of freedom are also interesting, defined as the energy and entropy density normalized by their Stefan-Boltzmann limit.

\begin{equation}\label{eq:degreesOfFreedom}
    g_\varepsilon (T) = \varepsilon(T)\frac{30}{\pi^2T^4} \quad , \quad g_s (T) = s(T) \frac{45}{2\pi^2T^3} \,.
\end{equation}

As stated above, we consider the various particle species in the Universe in thermal equilibrium. Furthermore, in the QGP phase, strong interactions dominate over other forces. Thus, potential electromagnetic/weak interactions between different particle species are negligible compared to strong interactions, which allows us to write thermodynamic quantities as the sum of the QGP contribution and the rest of the partial components from other sectors:

\begin{equation}\label{eq:cosmicThermodynamics}
    X_{\rm cosmic}(T) =  \sum_{i} X_i(T) \,,
\end{equation}
with $X \in {p, s, \varepsilon}$ and $i$ spanning the particle species. 
Let us list the particle species to consider in our temperature range $T\in [1;1300]$ MeV, with particle masses taken from \cite{ParticleDataGroup:2022pth}: 
\begin{itemize}
    \item 
    Quarks and gluons. Light quarks with flavors $u,d, s$  are included in the  microscopical model \cite{Blaschke:2023pqd} for the QCD EoS. The charm quark flavor $c$ with $m_c =1270$ MeV should be taken into account since $m_c \in [1;1300]$ MeV. We also consider the contribution of the bottom quark $b$ with $m_b=4180$ MeV. We recognize that the $b$ quark mass is above our temperature range and it already started to annihilate, but $b-\bar{b}$ annihilation is not instantaneous, and $80\%$ of particles annihilates in the temperature range $m_b>T>m_b/6 \sim 700$ MeV. Given the heavy mass of the top quark $t$ with $m_t=172.69$ GeV, we consider its contribution negligible in our temperature range. 
    \item Photons as pure radiation decouple from matter when the Universe becomes electrically neutral at $z\sim 1100$ \cite{WMAP:2008ydk}  with $T_{\rm CMB}(z) =2.728(1+z)$ Kelvin, giving $T_\gamma=0.26$ eV.
    \item Leptons: splitting the species into charged leptons ($e, \mu, \tau$) and neutrinos ($\nu_{e}, \nu_\mu, \nu_\tau$), gives two contributions. Neutrinos are considered massless particles and can thus be considered as pure radiation, they decouple at $T \simeq 1$ MeV and thus still interact in our temperature range. Charged leptons are considered to be an ideal fermion gas. 
    \item Non-strong bosons $W^\pm, Z^0, H^0$ are introduced as ideal boson gas, with respective masses $m_W =80.377$ GeV, $m_Z= 91.1876$ GeV, and $m_H = 125.25$ GeV. We recognize that their contribution should be close to zero in our temperature range. 
\end{itemize}

Using \eqref{eq:cosmicThermodynamics}, let us explicitly write :

\begin{equation}
    X_{\rm cosmic}^{(n_f)} = X^{(n_f)}_{\rm QGP} + X_\gamma + X_\nu + X_{e,\mu,\tau} + X_{\rm bosons}
\end{equation}
with the superscript $(n_f) = (2+1) , (2+1+1), (2+1+1+1)$ symbolizing the cases where the heaviest quark flavor in the cosmic soup is the strange, the charm, or the bottom quark, respectively. Details on the calculation can be found in the Appendix \ref{app:CosmicEoS}. Figure \ref{fig:PartialPressure} shows the partial pressure contributions.

\begin{figure}[t]
    \centering
    \includegraphics[width=0.95\linewidth]{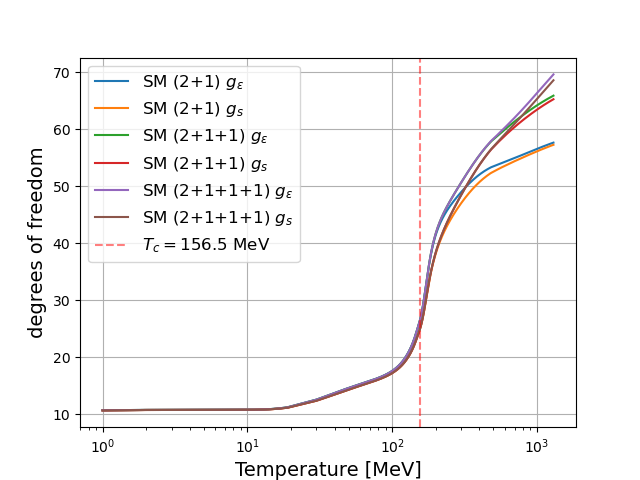} 
    \caption{Temperature dependence of the effective number of relativistic degrees of freedom.}
    \label{fig:Degrees of freedom}
\end{figure}

\begin{figure}[t]
    \centering
    \includegraphics[width=0.95\linewidth]{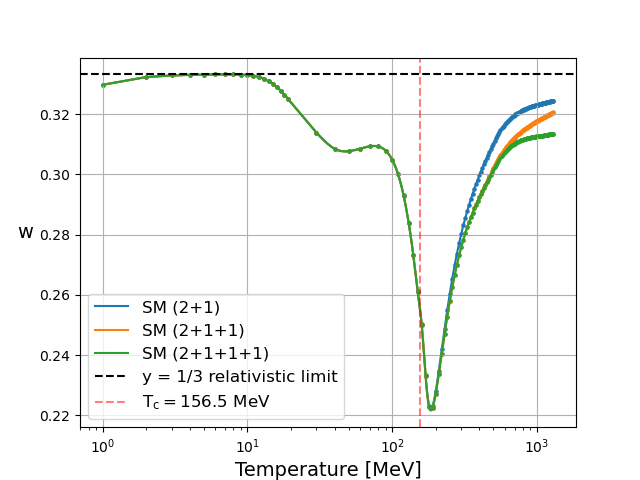} 
    \caption{Cosmic EoS for different numbers of quark flavors considered; the interpolated data is seen as solid line while the original data are explicitly shown as points. The black dashed line corresponds to the relativistic limit $w=1/3$.}
    \label{fig:CosmicEoS}
\end{figure}

We show $g_\varepsilon$ and $g_s$ in figure \ref{fig:Degrees of freedom}, the EoS is displayed in figure \ref{fig:CosmicEoS}. 
Both figures clearly show the effect of the cosmic phase transition on thermodynamics, and we see a clear dip in EoS around the QCD transition followed by a shoulder due to pion excitation before returning to the relativistic value $w=1/3$, as pure radiation dominates again (Figure \ref{fig:PartialPressure}). 
At low temperature, we see $w$ dipping again, which corresponds to $e^-e^+$ annihilation at $T\sim m_e$. 
In the same figure, we demonstrate the importance of accounting for different quark flavors which appear sequentially according to their masses. 
We even observe the influence of the heavy $b$ quark, since the (2+1 + 1) scenario deviates from the (2+1 + 1 + 1) scenario at $T\gtrsim m_b/6\sim 700$ MeV. 
Such results have been produced in the scientific literature using QCD lattice data \cite{Borsanyi:2016ksw,Byrnes:2018clq,Carr:2019kxo,Bodeker:2020stj,Escriva:2022bwe,Musco:2023dak} or phenomenological estimates from \cite{Laine:2006cp}.

\subsection{Beyond the Standard Model}
\label{ssec:BSM}
While the SM is a very successful mathematical description of matter and its interactions (except gravity), it cannot be the ultimate fundamental theory of nature. 
Indeed, many extensions, denoted as beyond the SM (BSM) physics, have been proposed; the best example would probably come from cosmology, where SM fails to provide a theoretical framework for DM. 
Even on a particle scale the SM fails to explain the nonvanishing masses of neutrinos. 
The paradigm of modern cosmology already uses several BSM physics processes \cite{Khlopov:2021xnw}. 
Experimental research on BSM physics is a particularly active field; so far, many "anomalies" (most of the time in the form of resonances) have been found in nuclear and particle physics experiments. 
The extension in particle physics can occur in the sectors of fermions (spin 1/2), scalar bosons (spin 0), and vector bosons (spin 1). 
Crivellin and Mellado recently published a review of anomalies in particle physics \cite{Crivellin:2025txc}.

A particularly significant deviation from the SM is the putative X17 scalar boson found in nuclear reaction $^{7}$Li$ (p, e^+e^-)^8$Be \cite{Krasznahorkay:2015iga} as well as in decay of excited nuclei $^4$ He and $^{12}$ C \cite{Krasznahorkay:2021joi,Krasznahorkay:2022pxs}. 
Recently, this resonance has also been found in the Positron Annihilation Dark Matter Experiment (PADME) \cite{2025arXiv250411439A}. 
References \cite{Krasznahorky:2024adr,Alves:2023ree} provide an overview of the studies on X17. 
As its name suggests, the X17 has a mass of $\sim 17$ MeV and is seen at the $6\sigma$ level \cite{Alves:2023ree} in all decay modes. 

\begin{figure}[t]
    \centering
    \includegraphics[width=0.95\linewidth]{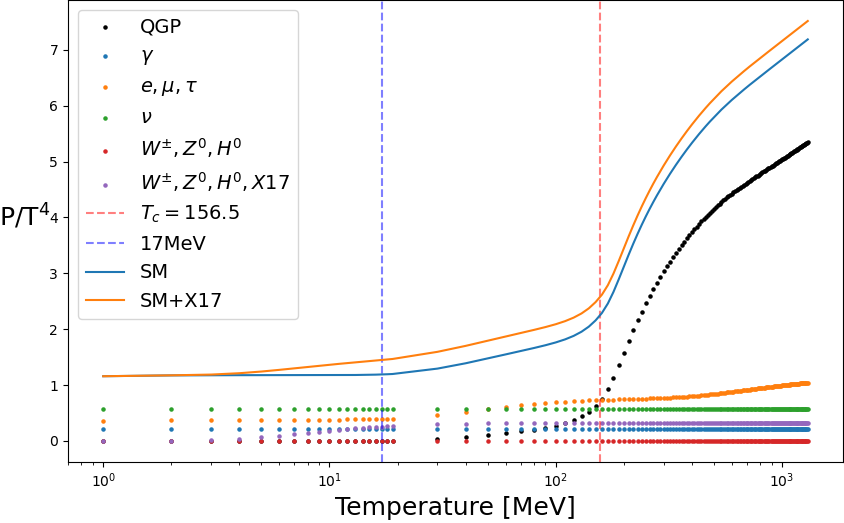} 
    \caption{Cosmic fluid pressure in the SM (2+1+1+1) and SM+X17 (2+1+1+1) case represented as continuous lines and their partial contribution represented with colored dots. The QGP pressure is bottom corrected, see appendix \ref{app:CosmicEoS}.}
    \label{fig:PartialPressure}
\end{figure}

In this section, we explore the effect of such a hypothetical particle on the cosmic EoS. We added X17 as a scalar boson $m_{\rm X17}=17$ MeV. Because of its low mass compared to SM bosons, the bosonic thermodynamics  in the SM+X17 case have a significant contribution to the cosmic thermodynamics in our temperature range. Figure \ref{fig:PartialPressure} highlights the partial contribution of X17 to the pressure with the orange dots and the total cosmic pressure with X17 is given by the orange continuous line. We observe the onset of the X17 contribution at a temperature of $T_{\rm X17}\sim m_{\rm X17}/6 \sim 3$ MeV while the onset temperatures of the lightest massive SM particles (muons and pions) are indistinguishably close at $T_\mu \sim m_\mu/6 = 17.6$ MeV and $T_\pi \sim m_\pi/6 = 23.3$ MeV, respectively.

\begin{figure}[t]
    \centering
    \includegraphics[width=0.95\linewidth]{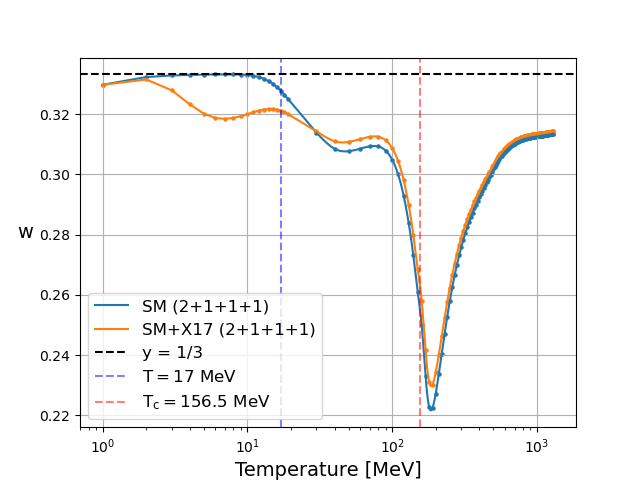} 
    \caption{Cosmic EoS in the SM (2+1+1+1) case and SM+X17 (2+1+1+1), the interpolated data is seen as a solid line while the original data are explicitly shown as points. The black dashed line corresponds to the relativistic limit $w=1/3$.} 
    \label{fig:X17CosmicEoS}
\end{figure}

In figure \ref{fig:X17CosmicEoS} we clearly see the impact an additional particle would have on the cosmic EoS, it not only adds a dip in $w$ at $T\sim m_{\rm X17}$, but also impacts the higher temperature behavior. The QCD dip and the pion shoulder are slightly mitigated in the presence of X17. The plot shows that no resonance can be ignored as we see that X17 has a noticeable impact on the overall cosmic EoS.  
In the temperature range below 1 MeV, which we do not consider in this work, no X17 effects are expected following the statement made in Subsection \ref{ssec:CosmicEoS}, because this temperature range is well below the threshold $T_{\rm X17}\sim m_{\rm X17}/6 \sim 2.8$ MeV. Consequently, in figure \ref{fig:X17CosmicEoS} the cosmic EoS of SM + X17 and SM become indistinguishable below $T\sim 2$ MeV, where the EoS dip again due to the annihilation of $e^+e^-$.
Moreover, since the X17 particle has a similar effect on the QCD peak as the inclusion of lepton asymmetry \cite{Bodeker:2020stj}, future studies should explore this. 

\section{PBH formation at the QCD transition 
}
\label{sec:pbh}

Fluctuations of energy density in the very early Universe are necessary to explain the current energy distribution of the Universe. Hawking and Carr introduced the idea that some of these early fluctuations could have been large enough to collapse to PBHs when the particle horizon crosses the fluctuation radius at $t_{\rm cross}$ \cite{Carr:1974nx}. Although early energy density fluctuations may have many possible origins, the main scenario is cosmic inflation, which typically assumes a Gaussian fluctuation spectrum. 
No matter what the source of the fluctuations is, they should be larger than the Jeans length at maximum expansion if they are to undergo gravitational collapse. Recently, Carr explored different formation scenarios \cite{Carr:2020xqk,Carr:2020gox}, and various studies provide numerical simulation of the collapse of overdensities \cite{Escriva:2022bwe,Musco:2023dak,Escriva:2019nsa,Musco:2004ak,Musco:2012au}. We refer interested readers to the 2021 review by Escriva \cite{Escriva:2021aeh} and the references therein.
Collapse can occur at different epochs during the radiation-dominated era, depending on the size of the fluctuation. Following Equation 4 in Bödeker et al. \cite{Bodeker:2020stj}, the particle horizon grows as follows:
\begin{equation}
\label{eq:horizonMass}
    T \approx 700~g_\varepsilon^{-1/4} \sqrt{M_\odot/M_H}~{\rm MeV}
\end{equation}
with $M_H$ the horizon mass.
The size of candidate PBH fluctuations increases with time; a full PBH mass spectrum is possible from Planck's mass ($10^{-5}$ g) if formed at Planck's time ($10^{-43}$ s) to supermassive range ($10^5 M_\odot$) for those formed as late as $1$ s after Big Bang \cite{Carr:2020gox}. 
Eq. \eqref{eq:horizonMass} translates our temperature range into the horizon mass range $M\in [3.4\times10^{-2}, 1.5\times10^5] M_\odot$. 
It should be noted that Eq. \eqref{eq:horizonMass} depends on $g_\varepsilon$, which is a thermodynamic function, so the horizon mass differs slightly between the SM and SM+X17 scenarios. The difference between the 2 horizon masses can be as high as 10\%, depending on the epoch; see Appendix \ref{app:horizonMass} for details.

\begin{figure}[t]
    \centering
    \includegraphics[width=0.95\linewidth]{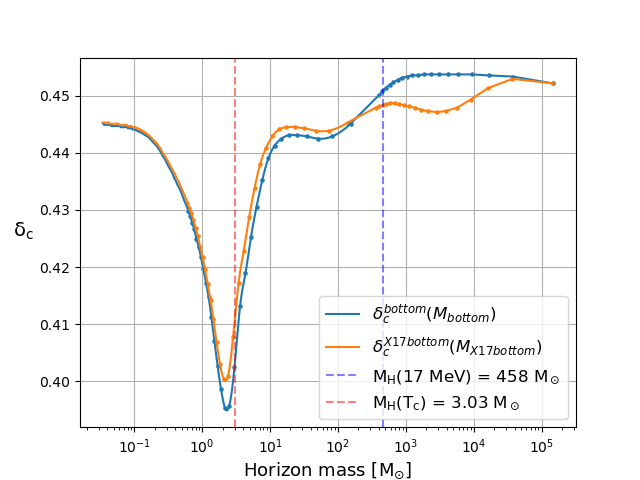} 
    \caption{Critical density threshold $\delta_c$ in the SM (2+1+1+1) case and SM+X17 (2+1+1+1). The interpolated data are shown as solid line while the original data are explicitly shown as points.}
    \label{fig:criticalDensity}
\end{figure}

Recent developments in lattice QCD simulations have allowed us to deepen our understanding of strongly interacting matter. In Ref.~\cite{Borsanyi:2016ksw}, it is shown that early phase transitions lead to drops in the number of relativistic degrees of freedom and dips in EoS values. We reproduced these results in Section \ref{ssec:CosmicEoS} using a microscopical model rather than lattice QCD. This framework allows us to retrieve QGP thermodynamics while having a phenomenological description of the QCD transition. The discrepancies in the EoS values of an ideal radiation fluid $w = 1/3$ are shown to affect the probability of collapse of a given overdense region and the induced PBH mass spectrum \cite{Byrnes:2018clq,Carr:2019kxo,Carr:2020gox,Bodeker:2020stj,Escriva:2022bwe,Musco:2023dak}. To comprehend how EoS can impact the probability of collapse and, consequently, PBH formation, we follow Carr's prescription from \cite{Carr:1975qj} to discriminate between collapsing and non-collapsing regions through a critical density contrast. One can provide a general definition of the threshold $\delta_c = \delta \varepsilon/\varepsilon$, where $\delta \varepsilon$ the energy overdensity and $\varepsilon$ the Universe's mean energy density. The value of $\delta_c$ depends on the initial fluctuation spectrum and the EoS. Such dependencies are determined by Musco \& Miller in \cite{Musco:2012au}, while Carr in \cite{Carr:2020gox} and the references therein provide an overview of the range of values of $\delta_c$. Using Figure 8 of \cite{Musco:2012au}, we show $\delta_c$ as a function of cosmic $w(M_H)$ in Figure \ref{fig:criticalDensity}. As expected, we observe the same features already seen in the EoS figure \ref{fig:X17CosmicEoS}. 

Based on the assumption of Gaussian inhomogeneities, the fraction $\beta$ of horizon patches undergoing gravitational collapse to PBH is:
\begin{equation}\label{eq:beta}
    \beta(M)\approx \mathrm{erfc}\left[\frac{\delta_c(w(T(M))}{\sqrt{2}\delta_{\rm rms}(M)}\right]
\end{equation}
where $M$ is the PBH mass, $\rm erfc$ is the complementary error function and $\delta_{\rm rms}$ is the root mean square amplitude of Gaussian fluctuation. Following \cite{Bodeker:2020stj,Carr:2019kxo}:
\begin{equation}\label{eq:delta_rms}
    \delta_{\rm rms} = A\times(M/M_\odot)^{(1-n_s)/4},
\end{equation}
where $n_s=0.97$ is the spectral index taken at its CMB value \cite{Planck:2018vyg}. The amplitude $A$ is a normalization parameter that expresses the strength of the fluctuations.

\begin{figure*}[t]
    \centering
    \includegraphics[width=\textwidth]{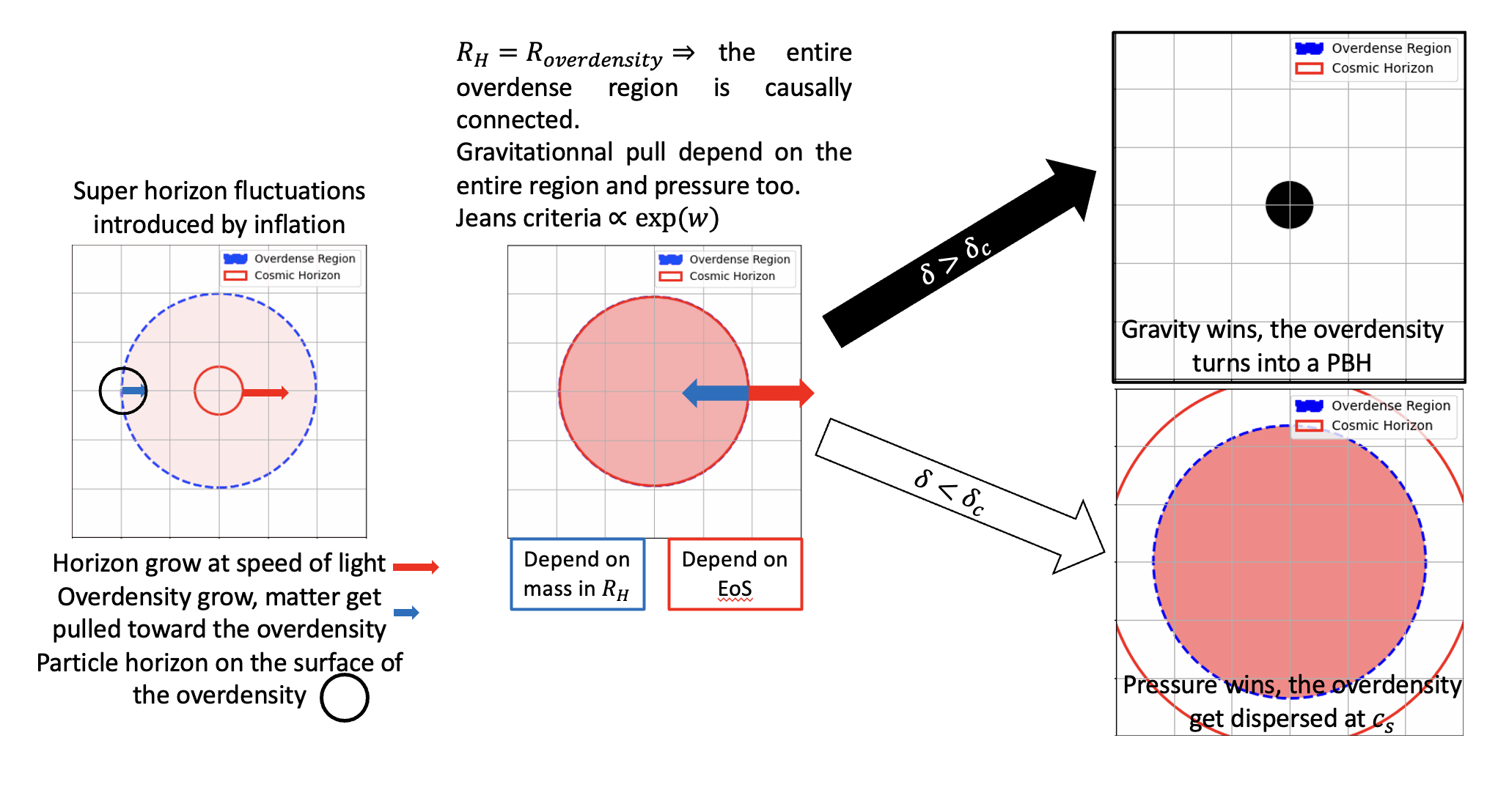} 
    \caption{Schematic view of possible paths followed by early energy density fluctuations. The left panel corresponds to an arbitrary initial time $t_i$ with $t_{\rm inflation} < t_i < t_{\rm cross}$; the middle panel corresponds to $t=t_{\rm cross}$. The right panels represent the fate of the fluctuations at a time $t>t_{\rm cross}$. An animated GIF version of the figure can be shared on demand: mael.gonin@dzastro.de} 
    \label{fig:pbh_formation}
\end{figure*}

The present fraction of DM in PBH of mass $M$ is then:
\begin{equation}
\frac{df_{\rm PBH}(M)}{d \ln M} \approx 2.4 ~\beta(M)\sqrt{M_{\rm eq}/M}
\end{equation}
where $M_{\rm eq}$ is the horizon mass at matter-radiation equality, the numerical factor $\rm 2.4=2(1+\Omega_b/\Omega_{\rm CDM})$, with $\rm \Omega_b=0.0456$ and $\rm \Omega_{\rm CDM}=0.245$ being the baryon and CDM density parameters from \cite{Planck:2018vyg}. 

\begin{equation}\label{eq:f_pbh}
    f_{\rm PBH} \equiv \int_{M_{\rm min}}^{M_{\rm max}} \frac{df_{\rm PBH}}{d \ln M} \,d \ln M\,.
\end{equation}
In a first approximation, the dynamics of PBH formation can be understood solely with the $\delta_c$ parameter. Fig. \ref{fig:pbh_formation} provides a schematic overview of PBH formation from inflationary fluctuations. We did not represent the detailed dynamics of the collapse, reference \cite{Carr:2019hud} studied such processes and their potential link with baryogenesis.

\begin{figure}[t]
    \centering
    \includegraphics[width=\linewidth]{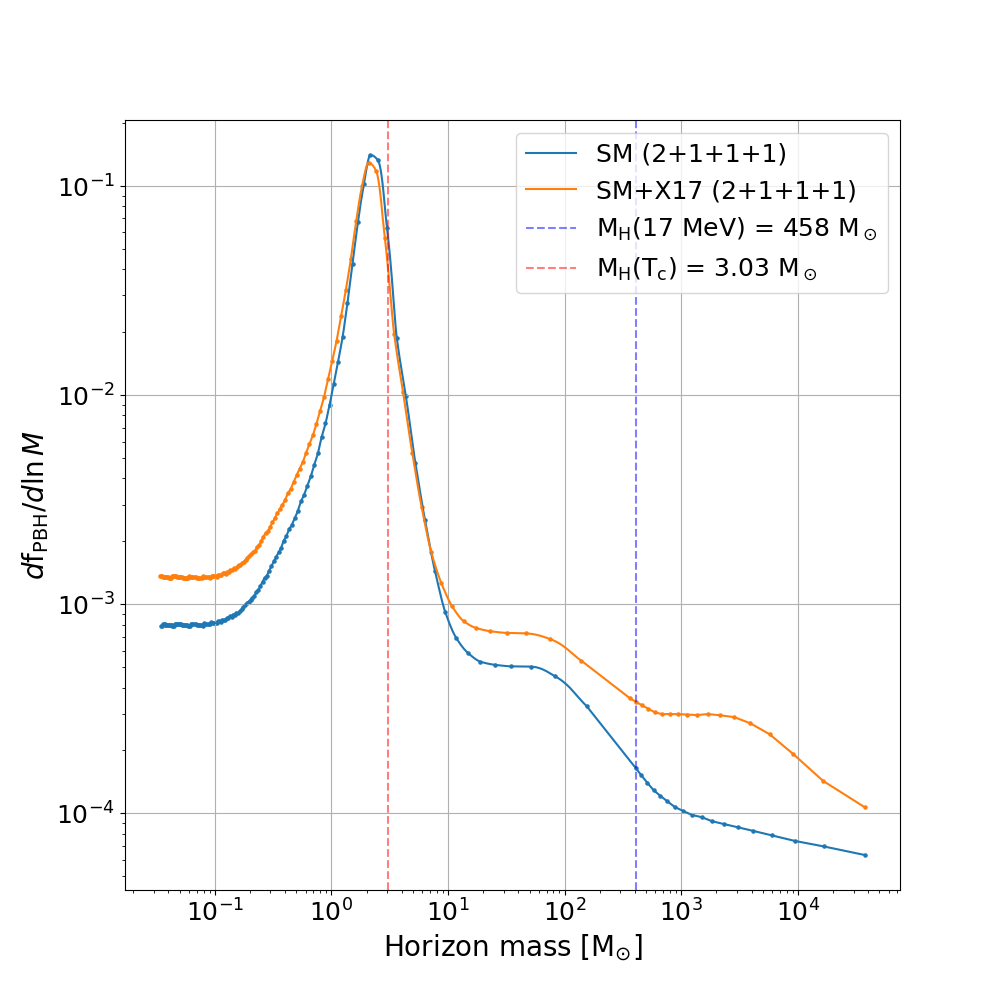} 
    \caption{PBH distribution for $f_{\rm PBH}=0.1$ in the SM and SM+X17 scenario.}
    \label{fig:pbhDistribution}
\end{figure}

Although the PBHs mass spectrum remains uncertain and dependent on many parameters, such as the fluctuation spectrum \cite{Carr:1974nx}, or lepton flavor asymmetry \cite{Bodeker:2020stj}, these results on early phase transitions allow one to determine the main features of the PBH mass spectrum as seen in Figure \ref{fig:pbhDistribution}. We again retrieve the main features of $w(T)$ and $\delta_c$ with a large peak at QCD transition $\sim 2M_\odot$ followed by the pion shoulder $M\in[10:100]M_\odot$. We also observe the effect of X17 on the PBH distribution, forming an additional shoulder at heavy horizon masses $M>10^3M_\odot$, while shifting the distribution upward everywhere except at the QCD transition peak, where it is slightly mitigated. We discuss this behavior later in this section.

In both scenarios, we observe the QCD peak spanning from $\sim 10\, M_\odot$ to $\sim 0.1\,M_\odot$, thus forming a substantial amount of PBHs below $1\,M_\odot$. This feature is of great importance since no astrophysical BH can be smaller than $\sim 5 M_{\odot}$, and neutron stars have a mass of $\sim1.4\,M_\odot$. While compact object candidates exist in this theoretical mass gap, their nature remains uncertain (they could be heavy neutron stars or exotic quark stars). Detection of BHs below $5\,M_\odot$ would strongly point to a primordial origin; furthermore, a firmly identified BH with $\sim 1\, M_\odot$ would provide smoking gun evidence for PBHs. The mass gap region is represented in Figure \ref{fig:GWhistogram}.

In the SM+X17 scenario, the formation of Intermediate Mass Primordial Black Holes with $100<M<10^5M_\odot$ is enhanced. 
Hereafter, we refer to them as IMPBHs. 
While there are many observations of stellar mass BHs and Super Massive Black Holes (SMBHs), there are still no direct observations of IMBHs, although there are also no constraints against their existence \cite{Greene:2019vlv}. 

Since the launch of JWST, evidence for a population of SMBHs in the early Universe has accumulated. These SMBHs are too heavy to be explained by formation through stellar evolution followed by Eddington rate accretion. To explain such massive BH populations, two explanations are viable but remain unproven. High-redshift SMBHs require heavy seeding \cite{Pacucci:2023oci} or super-Eddington accretion rates \cite{Schneider:2023xxr}, or a combination of both. These IMPBHs could provide seeds for high-redshift SMBHs \cite{Goulding:2023gqa,CEERSTeam:2023qgy}. PBHs formed during $e^-e^+$ annihilation with $M\sim 10^5-10^6 M_\odot $ are also heavy seed candidates for high-redshift JWST SMBHs \cite{Carr:2020gox}. Another heavy seeding scenario involves the direct collapse of gas clouds to SMBHs \cite{2020MNRAS.492.4917L}. Population III stars and their remnant BHs, while still not observed, could also provide seeds \cite{Nakazato:2005ek}. Suh et al. \cite{Suh:2024jbx} find evidence for super-Eddington accretion at high redshift. While Trinca et al. \cite{2024arXiv241214248T} find that episodic super-Eddington accretion rate triggered by galaxy merger is one explanation for the overmassive BHs at high redshift. Volonteri et al. \cite{Volonteri:2021sfo} argue that the accretion growth of seeds with masses $10^2-10^4 M_\odot$ is stunted due to their small gravitational sphere of influence compared to galactic scales.

Given the lack of consensus on IMBH accretion we argue that X17 IMPBHs, while lying at the low end of heavy seeds (see Figure 5 in \cite{Pacucci:2023oci}) represent viable candidates for high-redshift SMBHs when considering super-Eddington accretion.

We list here some of the interesting conclusions from 2020 IMBH review \cite{Greene:2019vlv}. Evidence for heavy IMBHs with masses $10^5-10^6 M_\odot$ in small galaxies is accumulating, and the X17 peak could provide seeds for such objects through early effective accretion. Moreover, cases can be made for IMBHs in some globular clusters. Wandering IMBHs are also expected in the galactic halo due to gravitational runaway in star cluster or those formed in satellite galaxies. 
Although the SM EoS already produces candidate IMPBHs, we argue that the SM+X17 scenario improves the probability of a primordial origin for such objects. It is worth noting that the primordial origin for IMBHs represents only one possible formation channel. Population III star remnants and direct collapse of early gas clouds are among the candidate mechanisms. 

Additionally, as stated in Section \ref{ssec:BSM}, the X17 boson is expected to have a negligible cosmological impact at low temperatures $T \lesssim m_{\rm X17}/6$, as well as before the QCD transition, in the high temperature regime. This behavior is indeed observed, with the two scenarios converging around $T \sim m_{\rm X17}/6$, as shown by the evolution of $w(T)$ in Figure \ref{fig:CosmicEoS} and $\delta_c$ in Figure \ref{fig:criticalDensity}.
However, the PBH mass distribution presented in Figure \ref{fig:pbhDistribution} does not indicate convergence of SM+X17 with the SM predictions.  
Further differences are observed. In addition to the difference at the QCD peak, SM+X17 consistently produces higher values than the SM leading to an absence of intersection between SM+X17 and SM at $\sim 200M_\odot$ as observed in Figure \ref{fig:criticalDensity}. This overproduction appears larger than it actually is due to the logarithmic scale, see Table \ref{tab:PBHmassFractions}.

These changes emerge during normalization. The process is carried out as follows: we choose a target $f_{\rm PBH} = 0.1$, then use a root-finding algorithm to determine the optimal value $\rm A_{opt}$ in Equation \ref{eq:delta_rms}, and subsequently compute $\rm \frac{df_{\rm PBH}}{dlnM}$ using $\rm A_{opt}$. We find a small but significant relative difference of approximately $1\%$ between the values of $\rm A_{opt}$ for SM and SM + X17, which means that for the same target $f_{\rm PBH} = 0.1$, SM+X17 requires stronger fluctuations: $\rm A_{opt}^{SM+X17}>A_{opt}^{SM}$. We explain this by the fact that X17 begins to contribute significantly to the pressure at the QCD transition (see Figure \ref{fig:PartialPressure}), creating a harder EoS. Hence, a region requires a higher critical density threshold value to collapse when X17 is present compared to the SM. Since most PBHs are located in the QCD peak (see Table \ref{tab:PBHmassFractions}), $\rm A_{opt}^{SM+X17}$ primarily compensates for the difference between SM and SM+X17 during the QCD transition: $\delta_w(\sim T_c) = w_{SM+X17}(\sim T_c) -w_{SM}(\sim T_c)$. It follows that regions where the difference between the two EoS values is smaller than $\delta_w(\sim T_c)$ form more PBHs in SM+X17 than in the SM. Using the same $A_{opt}$ for both scenarios, the exact relationship between SM and SM+X17 seen in Figures \ref{fig:criticalDensity} and \ref{fig:CosmicEoS} is retrieved, but the induced $f_{\rm PBH}$ differs; see Appendix \ref{app:PBHdistribution}.

Although the normalization in Figure \ref{fig:pbhDistribution} appears different, both models are strictly normalized to account for $10\%$ ($f_{\rm PBH} = 0.1$) of the total dark matter density. In fact, the relative overproduction of light, pion-scale, and intermediate-mass PBHs in the SM+X17 scenario is compensated by a slight but significant suppression of the QCD peak, as shown in Table \ref{tab:PBHmassFractions}. As expected, the most pronounced enhancement occurs in the X17 mass range, where the PBH fraction increases by nearly a factor of three compared to the SM. This is followed by the ‘lite IMBH’ regime, where the fraction almost doubles. Similarly, the low-mass and ‘pion shoulder’ regimes exhibit increases of approximately a factor of 1.5. The QCD peak represents the only region that experiences a decrease in PBH abundance, with the suppressed fraction redistributed toward the low-mass range and high-temperature regime well before the onset of the X17 transition.

\begin{table*}[t]
\centering
\begin{tabular}{|l|c|c|}
\hline
\textbf{Mass Range [$M_\odot$]} & \textbf{Standard Model} & \textbf{Standard Model + X17} \\
\hline 
Low mass range [$3.4\times10^{-2}$; 1] & 6.46\% & 10.07\% \\
\hline
QCD peak [1; 10]  & 91.53\% & 86.44\% \\
\hline
‘Pion shoulder’ [10; 100]  & 1.24\% & 1.76\% \\
\hline
‘Lite IMBH’ [100; 1000]  & 0.49\% & 0.91\% \\
\hline
X17 range [1000; $3.73\times10^4$]  & 0.28\% & 0.82\% \\
\hline
\end{tabular}
\caption{Fraction of total PBH abundance in different mass ranges for the Standard Model and Standard Model + X17 scenarios, considering only our limited temperature region.}
\label{tab:PBHmassFractions}
\end{table*}

\section{PBH as dark matter 
}
\label{sec:dm}
In this section, we provide a brief overview of PBHs as DM candidates within the mass range of interest. 
While this is by no means a detailed or comprehensive review, we highlight notable effects and implications of PBHs as DM, particularly in the intermediate-mass range where our SM+X17 scenario produces significant effects. For more thorough coverage, we refer interested readers to the works of Carr et al. \cite{Carr:2020xqk,Carr:2020gox,2024arXiv240605736C,Carr:2023tpt} and to a recent review by Bagui et al. \cite{LISACosmologyWorkingGroup:2023njw}.

PBHs represent compelling objects across a broad mass range through their fundamental connection to BH physics, as discussed in Section~\ref{sec:pbh}. These objects have already motivated Hawking to develop significant theoretical advances in BH research, unifying general relativity, quantum mechanics, and thermodynamics.
The concept of black hole evaporation emerged when Hawking examined the possibility of low-mass PBH formation before the QCD epoch \cite{2024arXiv240605736C,Carr:2023tpt}. 
Such objects would experience quantum effects, radiating particles like black bodies at temperature $T\sim M^{-1}$ and evaporating on a timescale $t_{\rm evap} \sim M^3$ \cite{Hawking:1975vcx,Hawking:1974rv}. 
PBHs also serve as robust cold DM candidates, as their nature makes them non-relativistic, non-baryonic, and nearly collisionless dark objects. Recent developments in GW astronomy, combined with the absence of pDM discoveries despite three decades of intensive experimental efforts, have renewed interest in PBH-DM scenarios. These models could potentially resolve several major puzzles in contemporary cosmology \cite{Carr:2019kxo,Carr:2023tpt}.

Since PBHs remain neither observationally confirmed nor definitively excluded, our understanding emerges from evaluating competing constraints against positive evidence. We organize this section into two subsections: Section \ref{subsec:constraints} examines constraints on PBH-DM models, while Section \ref{subsec:PositiveEvidence} discusses both the limitations of these constraints and the positive circumstantial evidence supporting PBH existence.

\subsection{Constraints}\label{subsec:constraints}
\begin{figure*}[t]
    \centering
    \includegraphics[width=\linewidth]{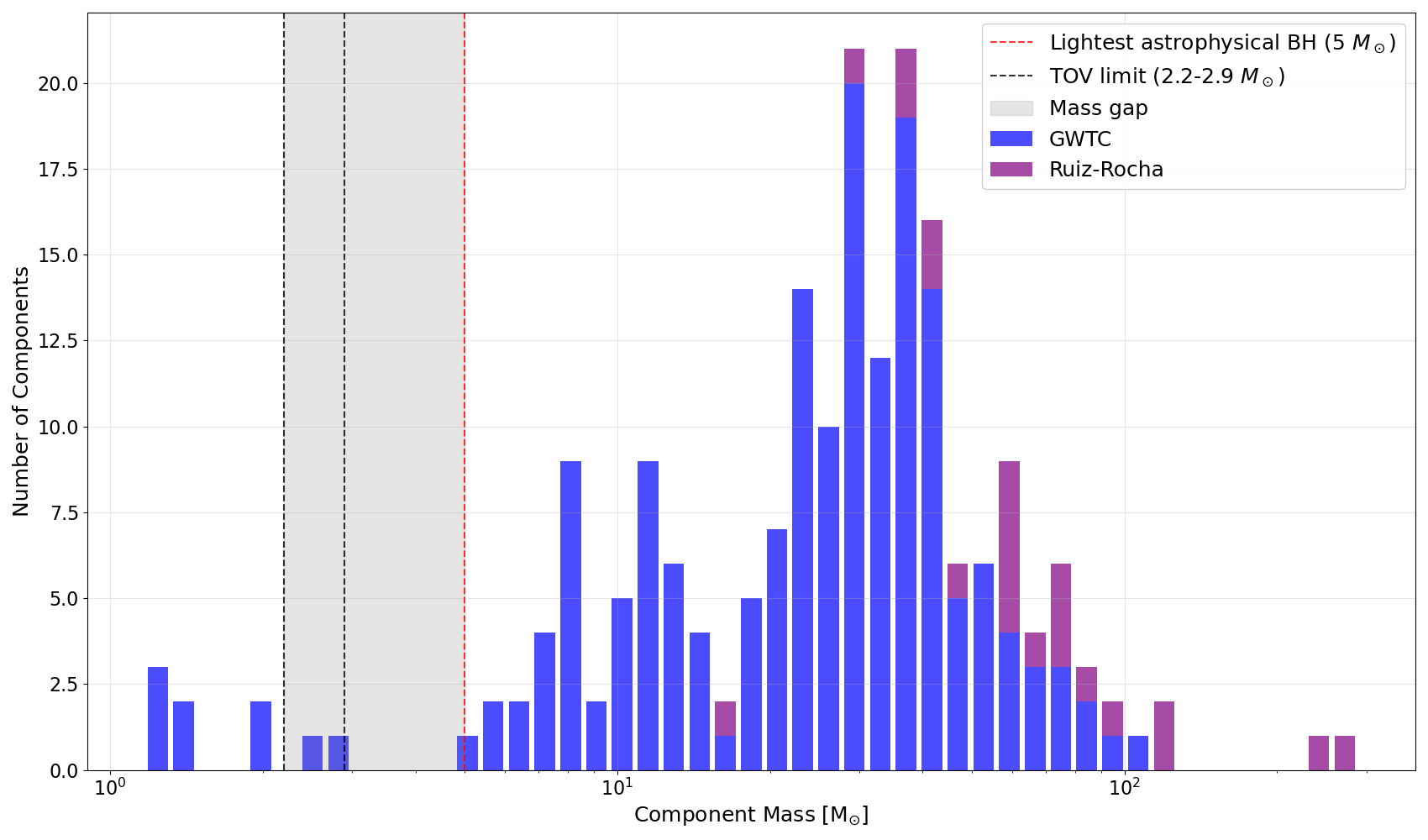} 
    \caption{Histogram of gravitational wave events sorted by mass and discriminated by origin. The purple colors correspond to the recent Ruiz-Rocha et al. \cite{2025arXiv250217681R} study. TOV = Tolman–Oppenheimer–Volkoff, corresponding to the theoretical limit on neutron star mass. The TOV limit depends on the still undetermined neutron star equation of state; this uncertainty is represented by the two dotted lower limits.}
    \label{fig:GWhistogram}
\end{figure*}

The proposal of Hawking radiation allows us to split PBHs into two categories: evaporating and non-evaporating PBHs. The former are sufficiently light BHs $M\lesssim 10^{-16}M_\odot\sim 10^{17}$ g, such that quantum effects are significant and they would have evaporated during the lifetime of the universe, while the latter are heavy enough to survive longer than the current cosmic age. Carr et al. \cite{Carr:2020gox} review constraints on evaporating PBHs, from Big Bang Nucleosynthesis (BBN), cosmic microwave background (CMB), extragalactic $\gamma$-ray background and galactic $\gamma$-ray background. Since evaporating PBHs fall outside our PBH mass range, we do not discuss this topic further. For the non-evaporating category $M\gtrsim 10^{-16}M_\odot$, we can distinguish two families of constraints: 
\begin{itemize}
    \item Direct observational constraints: Lensing and GW. Since PBHs are compact objects, they should interact as such, lensing luminous objects in their background and producing GW when orbiting in binary motion. 
    In our PBH mass range EROS, OGLE, and Icarus have produced constraints on the composition of Milky Way's halo, excluding planetary and stellar mass compact objects from comprising more than $\sim 10\%$ of the halo masses \cite{Wyrzykowski:2011tr,EROS-2:2006ryy,Oguri:2017ock}.
    GW limits are provided through successive catalogs \cite{LIGOScientific:2018mvr,LIGOScientific:2020ibl,KAGRA:2021vkt,KAGRA:2021duu}; see Figure \ref{fig:GWhistogram} for a mass histogram of GW events. Franciolini et al. \cite{Franciolini:2022} established an upper bound on PBH populations through Bayesian analysis of LIGO-VIRGO-KAGRA (LVK) data. Additionally, PBHs should contribute to the stochastic GW background through mergers and formation processes; Domènech recently reviewed these mechanisms \cite{2024arXiv240217388D}. 
    
    \item Indirect observational constraints: Dynamical and accretion. Phenomena such as dynamical heating are influenced by PBH-DM due to frequent PBH-star, PBH-stellar remnant BH, PBH-PBH encounters. This effect is particularly pronounced in highly DM dominated environments, where PBH-DM cannot be considered effectively collisionless, such as ultra-faint dwarf galaxies (UFDGs). Systems such as Eridanus II \cite{Brandt:2016aco,Green:2016xgy,DES:2016vji} and Segue I \cite{Koushiappas:2017chw} provide constraints within our mass range. The presence of an IMBH in the center of Eridanus II remains under debate; constraints on PBHs are relaxed if the central globular cluster of Eridanus II contains an IMBH \cite{DES:2016vji}. 
    PBHs should be accreting matter, thereby impacting the Universe's thermal history and X-ray background \cite{2024A&A...688A.183C}. In mixed DM scenarios, where PBHs constitute only a fraction of the DM, Serpico et al. \cite{Serpico:2020ehh} argue that IMPBHs would accrete dark matter into mini-halos, enhancing their gravitational potential and potentially serving as seeds for SMBHs. Yuan et al. \cite{Yuan:2023bvh} have confirmed this finding. Such mechanisms constraints IMPBHs abundance through their accretion feedback processes.
\end{itemize}

Regarding the dynamical constraints, while SM+X17 overproduces PBHs with $M\gtrsim10M_\odot$ compared to the SM scenario, making the model strongly constrained by UFDG dynamics, it also produces a population of IMBHs. Following the prescription of Li et al.\cite{DES:2016vji} for compact systems containing an IMBH, the X17 peak in the IMBH range could mitigate the dynamical constraints.
When we combine recent observational constraints on the PBH contribution to DM in the mass range 0.1-100 solar masses from OGLE microlensing \cite{Mroz:2024mse}, LVK searches for subsolar mass black holes \cite{LVK:2022ydq}, and accretion limits on the X-ray background contribution \cite{2024A&A...688A.183C}, the original predictions for the PBH abundance by Carr \cite{Carr:2023tpt} and Hasinger \cite{Hasinger:2020ptw} appear to be overestimated by approximately a factor of ten. In particular, the strong QCD peak around the Chandrasekhar mass appears to be ruled out.

\subsection{Discussing the constraints and positive evidences}\label{subsec:PositiveEvidence}

On the positivist side of PBH-DM research, several observational evidences and unexplained cosmological mysteries point to the existence of PBHs \cite{Carr:2023tpt,Clesse:2017bsw}. It is important to emphasise that all limits on the abundance of PBHs come with significant caveats. One particularly unrealistic assumption in most constraints is a monochromatic PBH mass spectrum, which is convenient for modeling, but unrealistic given the very early cosmic EoS presented in Section \ref{ssec:CosmicEoS} and the resulting PBH distribution in Section \ref{sec:pbh}. This extended mass function of PBHs allows PBHs to explain a significant fraction of DM \cite{Carr:2017jsz,Kuhnel:2017pwq}. Even if our mass range is constrained, we can still argue for the existence of the mass function here.

The constraints imposed by different observation channels each have important limitations. The constraints imposed by gravitational waves remain weak due to their strong model dependence, as several mechanisms can suppress the observable PBH merger rate. In a series of articles, Jedamzik \cite{Jedamzik:2020ypm,Jedamzik:2020omx} has shown, for example, how considering several mass bins of PBHs in clusters can lead to a suppression of the PBH merger rate. Furthermore, dynamic friction by ambient dark matter could lead to rapid mergers, with early-formed PBH binary systems possibly merging before $z\sim1$ \cite{Hayasaki:2009ug}. PBH could also trigger early structure formation through the Poisson and seed effects \cite{Carr:2018rid,Carr:2019bel,Delos:2024poq}. Future gravitational wave telescopes should provide better constraints through discoveries at high redshifts, measurements of the stochastic gravitational wave background, and mergers of sub-solar masses \cite{Franciolini:2023opt,Barsanti:2021ydd}.

The accretion constraints must be viewed with caution, as they are still highly model-dependent. Garcia-Bellido \& Hawkins \cite{Garcia-Bellido:2024yaz}, for example, have pointed out that the microlensing limits become significantly weaker when the current Gaia DR3 rotation curve is assumed instead of a flat rotation curve for the Milky Way. Carr et al. \cite{Carr:2020gox} argue that accretion constraints are weak compared to constraints from dynamic processes due to their high model dependence, as most accretion constraints assume Bondi-Hoyle-Littleton accretion \cite{Bondi:1952ni}, for which there is no observational evidence yet.
Furthermore, the accretion limits are based on heuristic models for the complicated accretion process, which are calibrated in the local Universe and not at high redshift, and typically assume monochromatic mass distributions. If PBHs have a broad mass distribution or form clusters, these limits become less stringent. Importantly, physics beyond the Standard Model, such as the X17 particle or the introduction of a high lepton flavour asymmetry \cite{Bodeker:2020stj}, can modify the predicted PBH mass spectrum so that it is less in conflict with observational constraints.

A mixed DM scenario should also be considered. In a series of papers, Agius et al. \cite{Agius:2024ecw} and Scarcella \cite{Scarcella:2024lco} have examined the Bondi-Hoyle-Littleton accretion assumption and instead used the Park \& Ricotti BH \cite{Park:2012cr} accretion model, which is supported by hydrodynamic simulations. They conclude that DM mini-halos should not amplify BH accretion as much as previously assumed, which significantly weakens Serpico's constraints \cite{Serpico:2020ehh} and confirms Carr's argument regarding accretion constraints.

Similar to the constraints, the arguments for PBHs can be divided into two families:
\begin{itemize}
    \item Direct evidence: lensing effect and gravitational waves. In the early 2000s, the MACHO (MAssive Compact Halo Objects) project discovered 17 microlensing events in the Large Magellanic Cloud (LMC), sparking a discussion about compact objects with a mass below the solar mass $M\sim 0.5 M_\odot$, which make up 20\% of the LMC's halo \cite{MACHO:2000qbb}. The origin of these events remains unknown, although it should be noted that such a population is consistent with the QCD peak. This 20\% limit is subject to uncertainties regarding the shape of the halo. Hawkins argues in \cite{Hawkins:2015uja} that the limit is not certain and that 100\% of DM could be contained in PBHs with solar mass.
 
 Since 2015 and the first GW detection \cite{LIGOScientific:2016aoc}, more than 90 GW signals have been confirmed, with the fourth observation campaign still ongoing since May 2023. PBHs are natural candidates for these GW signals involving BHs, especially BH-BH mergers. Recently, a reanalysis of the LVK O3 run revealed 11 candidates for BH-BH mergers in the ‘lite IMBH’ mass range $M \sim 10^2\, M_\odot$ \cite{2025arXiv250217681R}. We plot these events in Figure \ref{fig:GWhistogram}.

Comparing the BH-BH merger histogram in Fig. \ref{fig:GWhistogram} with the PBH distribution in Fig.~\ref{fig:pbhDistribution}, it can be argued that the pion shoulder at $\sim 30\,M_\odot$ can explain the bulk of the BH-BH merger population. Several studies have claimed that PBHs have indeed been observed by GW \cite{Bird:2016dcv,Sasaki:2016jop,Clesse:2016vqa}. 
    Under simple assumptions, the PBH mass distribution should predict a peak in GW detections around $1$ - $10\,M_\odot$, which corresponds to the QCD-induced peak; this behaviour is not clearly evident from the current data. 
However, there is clearly a BH population that lies there 
(Fig.~\ref{fig:GWhistogram}), with some objects in the theoretical mass gap between the lightest astrophysical BH, which originated from supernovae with about $\sim 5\,M_\odot$, and the heaviest neutron star with about $2.2$ to $2.9\,M_\odot$, depending on the neutron star's EoS considered. 
    Objects below the Tolman-Oppenheimer-Volkoff limit are interpreted as NS.
    Moreover, the observed excess abundance of $\sim 30\, M_\odot$ merger components could be explained using Jedamzik's conclusion \cite{Jedamzik:2020ypm,Jedamzik:2020omx} that lighter QCD PBH binary systems are more easily destroyed and therefore less likely to merge. 
    Furthermore, a mass gap of black holes formed by pair instability supernovae between 40 and 60 $M_\odot$ \cite{2019ApJ...887...53F} is expected; however, this gap is not observed in gravitational wave (GW) data. 
    The recent ‘Ruiz-Rocha’ black holes further populate this supposedly forbidden region. 
    The striking absence of such a mass gap suggests either new supernova physics or the existence of another population of black holes with different origins.
    
    In 2017, Raidal et al. \cite{Raidal:2017mfl} showed that a log-normal PBH distribution can explain the LIGO/Virgo events without violating the LIGO constraints. Raidal's conclusion must be viewed with caution, as they admit that for $f_{\rm PBH}>0.1$ in the solar mass range, their analytical prediction of the merger rate shows only a weak correlation with the results of  N-body simulations. Sten-Delos et al. confirmed this point in their 2024 study \cite{Delos:2024poq}.
Boedeker et al. \cite{Bodeker:2020stj} investigate how a large lepton asymmetry modifies the very early EoS and the subsequent PBH mass function. They find that the scenario with $l_e=-8\times 10^{-2}$ and $l_\mu = l_\tau= 4 \times 10^{-2}$, where $l_e, l_\mu$ and $l_\tau$ are the lepton asymmetries of electrons,  muons and tau leptons, respectively, is consistent with current BH-BH merger events, see Fig. 6 in \cite{Bodeker:2020stj}.
   
 \item Indirect circumstantial evidence: dynamics and accretion. UFDWs are used to limit the abundance of PBHs, but they can also provide clues about PBH DM. 
 There is an observed critical radius and critical mass for UFDGs.
 Clesse \& García-Bellido \cite{Clesse:2017bsw} argued that a population of MACHOs would dynamically heat the environment and prevent the survival of UFDGs below the observed critical radius between 10 and 20 pc.
 Furthermore, they point out that the large mass-to-light ratio of UFDGs can be explained by an early episode of rapid accretion that suppresses star formation. 
 With this framework, they also claim to be able to explain the lack of chemical evolution in UFDGs \cite{Brown:2012uq}. 
 Silk \cite{Silk:2017yai} has argued that IMBHs may have seeded dwarf galaxy AGNs in a gas-rich past, while they are inactive today. 
 Such a mechanism could solve a significant number of problems related to dwarf galaxies.   

   Thanks to the recently published GAIA DR3 data \cite{2021A&A...649A...1G}, Han et al. \cite{2025ApJ...982..188H} were able to trace 10 hypervelocity stars back to the LMC and rule out the possibility of a supernova runaway or dynamic ejection, discovering a dormant small SMBH $\sim 6 \times 10^5\, M_\odot$ just next to our galaxy. 
   Furthermore, using GAIA data, two dormant BHs $\sim 10\,M_\odot$ were found in binary star systems with stellar companions \cite{Chakrabarti:2022eyq,El-Badry:2022zih,2024MNRAS.527.4031T} in Gaia DR3 and another, more massive one $\sim 33\,M_\odot$ in Gaia DR4 pre-release \cite{Gaia:2024ggk}. 
    QCD and pion PBHs are natural candidates for these Gaia BHs, with the latter being of particular interest as the system is an escaper from the $\sim 13$ Gyr old ED-2 star cluster \cite{2024A&A...687L...3B}. 
    The discussion about its origin is lively, Ref. \cite{2024A&A...688L...2M} argues for a dynamic formation consistent with PBHs, while Ref. \cite{Iorio:2024pat} defends the idea of formation from an original binary star system consisting of a heavy ($40-60\,M_\odot$) metal-poor star and a low-mass star ($1\,M_\odot$).
  
    In their Nature article, Häberle et al. \cite{2024Natur.631..285H} claimed to have found a signature of $8200 M_\odot$ BH in the centre of  $\omega$ Centauri through observations of seven fast-moving stars in the central region of 0.08 pc.

Accretion could also have implications for early star formation \cite{Zhang:2023hyn,Cappelluti:2021usg}, but the model dependence is strong and different studies come to different conclusions. This could either be a constraint or a positive evidence for PBHs.
\end{itemize}

\begin{figure*}[!t]
    \centering
    \includegraphics[width=0.9\linewidth]{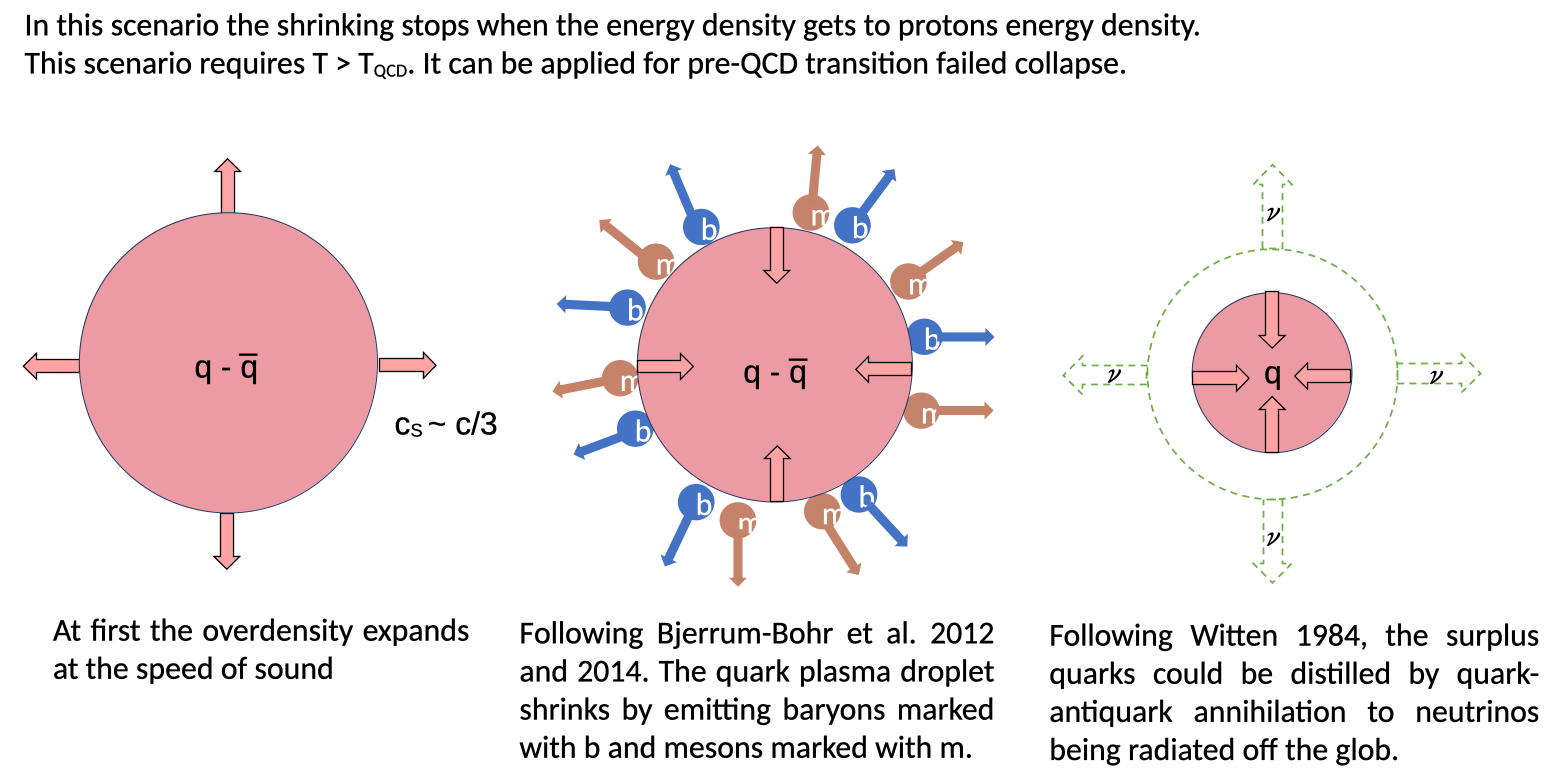}
    \caption{Evolution of a superdense region of finite size during the QCD phase transition, which does not transform into a PBH but nevertheless remains gravitationally bound long enough to form QGP globs. Left panel: Expansion and cooling at the speed of sound; Middle panel: Baryon number distillation through meson evaporation \cite{Barz:1990fs,Bjerrum-Bohr:2011gnk,Bjerrum-Bohr:2013qya}; Right panel: Enrichment of baryon density through quark-antiquark annihilation to neutrinos, which are emitted from the glob \cite{Witten:1984rs}.
    Animated GIF version of the figure can be shared on demand: mael.gonin@dzastro.de}
    \label{fig:failed-PBH}
\end{figure*}

\section{Primordial dense baryon matter clumps and heavy element
freeze-out 
}
\label{sec:elements}

The formation of primordial BH in the early Universe has strong
implications for models of Big Bang nucleosynthesis.
Standard models assume a homogeneous Big Bang nucleosynthesis (HBBN),
which occurs at the nucleation time $t_{\rm nucl}$ about 3 minutes after
the Big Bang.
Only H, He, and a small amount of Li are produced; all the other elements
are subsequently synthesized by burning processes or, in the case of the heavy
elements, by explosive events.
The existence of strong gravitational fields, for example due to
long-lived PBHs that survive $t_{\rm nucl}$ opens up alternative
scenarios of inhomogeneous Big Bang nucleosynthesis (IBBN).
Not all baryonic matter is in the low-density state of the HBBN at $t_{\rm nucl}$,
but dense, gravitationally bound clumps are formed, which survive the
nucleation time, so the formation of nuclei takes place under different
conditions.

As we have seen in section \ref{sec:pbh}, a particularly prominent era
for the formation of PBHs is the cosmic QCD transition
\cite{Musco:2023dak,Jedamzik:1999am}, which produces PBHs of about one solar mass. The success of the PBH scenario shows that the occurrence of
large fluctuations, which are even gravitationally supercritical, 
should be an integral part of an improved inhomogeneous Big Bang (IBB) scenario
\cite{Kajantie:1986hq,Ignatius:1993qn,Ignatius:1994fr}.
In such a scenario, in addition to PBHs, gravitationally subcritical
density fluctuations should also be present at the cosmic hadronization transition, which do not form black holes but rather dense quark-gluon
plasma globs of the corresponding Jeans mass. According to a scenario worked out for the situation in heavy ion collisions, these hadronize and can lead to the distillation of regions with high baryon number \cite{Barz:1990fs},
see the middle panel in Fig. \ref{fig:failed-PBH}.
For a recent update, see also
\cite{Bjerrum-Bohr:2011gnk,Bjerrum-Bohr:2013qya}.

In addition to the vaporization of hadronic shells from quark-gluon plasma globs
a second, very effective mechanism for the distillation of the baryon number
was proposed by Witten in his seminal work on the ``Cosmic separation
of phases'' \cite{Witten:1984rs}.
It consists in the annihilation of quark-antiquark pairs to neutrinos
which, due to their small cross section of interaction with matter,
have a longer mean free path than photons and can therefore cool the globs more effectively
by simultaneously enriching their net baryon density; see the right panel in Fig.
\ref{fig:failed-PBH}.
Following the Bodmer-Witten hypothesis that strange quark matter
is the absolutely stable ground state of high-density matter, it would
be attained at the end of the distillation process, with a net baryon
density of $n_B\sim 10^{15}$ g/cm$^3$. 
The size of these strange quark matter lumps was estimated by Witten to be less than $10^{18}$ g \cite{Witten:1984rs}, roughly equivalent to the mass of Mount Everest.
If the present-day baryon-to-photon ratio were a relic of antiquark annihilation in the cosmic hadronization transition, implying a baryon asymmetry at that epoch of the same order, $(n_q-n_{\bar q})/(n_q+n_{\bar q})\approx 10^{-9}$, then a subcritical density fluctuation (a failed PBH) should start the baryon distillation process with a mass around the solar mass. This would mean that dense quark matter globs should have a mass of up to $10^{-9}~M_\odot \approx 10^{24}$ g \cite{Boeckel:2009ej} at the end of the hadronization transition. This would correspond to the mass of the Ceres asteroid. With a density of $10^{15}$ g/cm$^3$, a spherical object with this mass would have a radius of around 6 meters.  
If one adopts the more conventional viewpoint that nuclear matter, rather than strange quark matter, represents the stable ground state, then Witten's neutrino-driven distillation scenario provides a mechanism for producing the gravitationally subcritical clumps of nuclear matter hypothesized above. Due to their low mass, these clumps of nuclear matter are not gravitationally stabilized and are subject to a Coulomb explosion that freezes out their chemical composition and scatters heavy $r$-process elements into the cosmos.

In this way, PBHs and the Witten mechanism represent two aspects of strong inhomogeneities at the cosmic QCD transition that have the potential to solve major cosmic puzzles: PBHs could account for the bulk of dark matter and explain the early formation of SMBHs as seeds of galaxies and quasars, while failed PBHs become sites of early formation of heavy $r$-process elements already in the 'dark ages', even before the birth of the first stars.

While the light elements up to the iron/nickel region are produced steadily in stellar burning processes, the heavier elements are mainly frozen out.
Special conditions are necessary to run the $r$ or $s$ process, which is recently possible, e.g., in SN explosions or NS mergers.
For a review on deciphering the origins of the elements through galactic archeology, see \cite{2025arXiv250318233F}, where further references can be found.
Heavy elements are observed in various astrophysical objects, but the site where they are formed has not been fully resolved yet.

\begin{figure}
     \centering
\includegraphics[width=\linewidth]{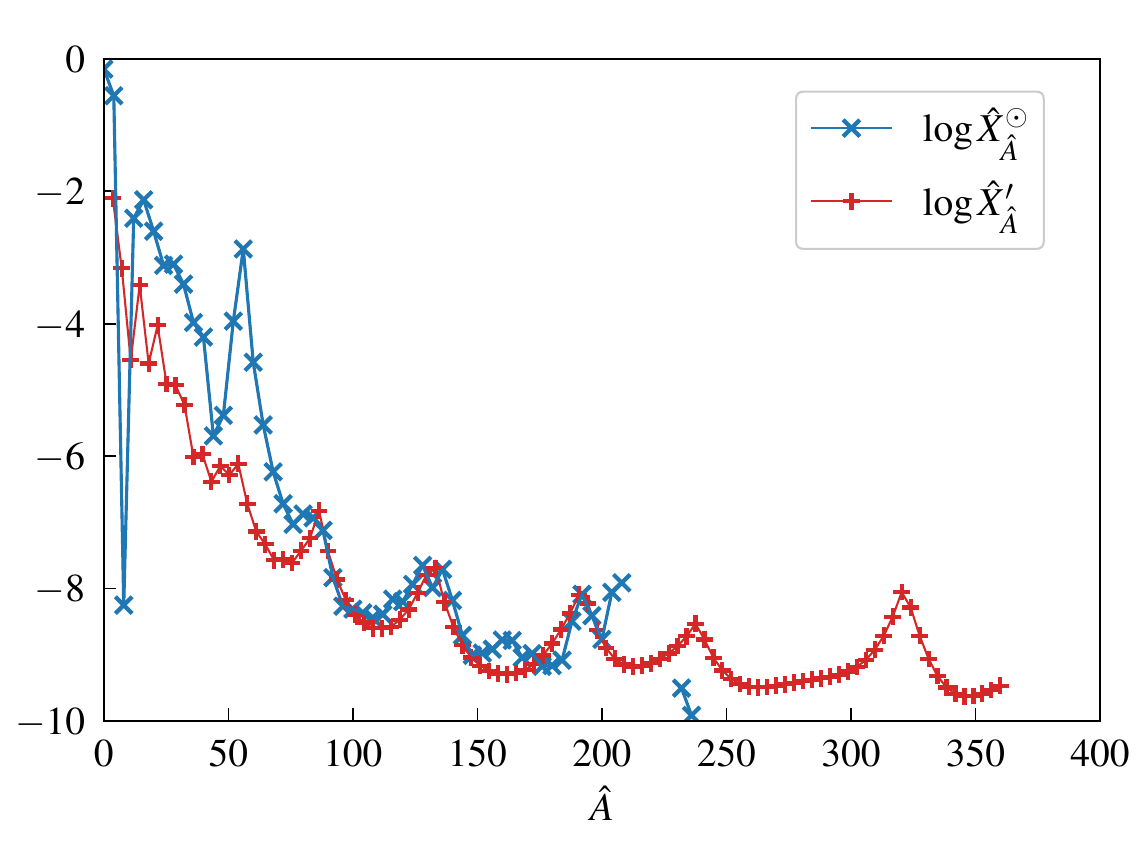}
\includegraphics[width=1.1\linewidth]{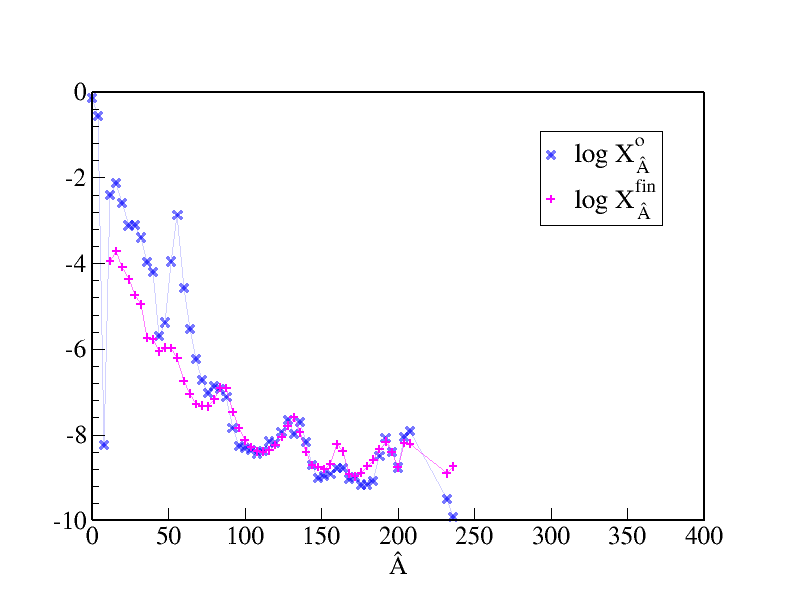}
     \caption{Upper panel: Accumulated mass fraction $\hat X_{\hat A}$ (red ``$+$''
symbols) for the parameter values $T=5.266\,\mathrm{MeV}$,
$\mu_n=940.317\,\mathrm{MeV}$ and  $\mu_p=845.069\,\mathrm{MeV}$ after
evaporation of neutrons $\hat X'_{\hat A}$ compared with the solar
accumulated mass fraction $\hat X^\odot_{\hat A}$ (blue ``$\times$''
symbols. For details, see text and Ref. \cite{2009LanB...4B..712L}). 
Lower panel: In
addition to neutron evaporation, leading to  $\hat X'_{\hat A}$, nuclei
with $A>212$ are subject to $\alpha$ decay (feeding the region near
$A\sim 200$) and fission (feeding the region near $A\sim 160$), 
with the final distribution $\hat X^{\rm fin}_{\hat A}$.
}
     \label{fig:XA}
\end{figure}

Recently, a sudden freeze-out scenario has been considered for the formation of heavy $r$-process elements \cite{2024arXiv241100535R}.
Lagrange parameters $\lambda_i$ are determined that are the non-equilibrium generalizations of temperature $T$ and of the chemical
potentials of neutrons $\mu_n$ and protons $\mu_p$. As an example, the accumulated mass fractions
\begin{equation}
\hat X_{\hat A}=\frac{1}{n_B} \sum_{A'=\hat A}^{\hat A+3} A'
\sum_{Z,\nu} n_{A'Z\nu}
\end{equation}
(with the baryon number density $n_B$,  $\nu$ denoting the excitation state 
of the isotope $\{A',Z\}$ with density $n_{A'Z\nu}$) for
the solar abundances are shown in Fig. \ref{fig:XA}.

The freeze-out distribution with the values of the Lagrange parameters $T=5.266 \,\mathrm{MeV}$, $\mu_n=940.317\,\mathrm{MeV}$, and
$\mu_p=845.069\,\mathrm{MeV}$ was calculated, and the accumulated mass
fractions $\hat X'_{\hat A}$ after neutron evaporation are also shown. 
We see a large number of superheavy elements that decay after
freeze-out. 
The $\alpha$ decay feeds down to the lead region explaining
the large amount in the final distribution, whereas fission feeds down to
elements in the region of rare earth elements. 
Since the branching ratios for the decay and fission of $\alpha$ for the superheavy extremal nuclei are not known, we can only give a simple estimate here
where the lowest mass numbers beyond $A=208$ perform $\alpha$ decay to reproduce the
observed values $\hat X_{204}, \hat X_{208}$, while the higher values of
$\hat A$ perform symmetric fission.  The result is shown in the lower panel of Fig. \ref{fig:XA}.
The enhancement in the range $A\approx. 160$ overestimates the solar data.
A better treatment of fission, taking into account asymmetric fission,
could smear out this peak. A detailed distribution of the isotopic abundances
would be obtained from reaction network calculations describing the
decay process after freeze-out in a microscopic way, but experimental data for the relevant nuclei are currently not available.

Another question is the site in the Universe where such conditions may occur. 
The standard answers are SN explosions, which, however, are not
sufficient to explain the formation of the heaviest $r$-process elements.
An alternative is NS mergers, which, indeed, may produce heavy elements.
However, the occurrence of heavy elements in the first objects is
also hardly to be understood in this way \cite{Thielemann:2020qmv,Wehmeyer:2020egj,Wehmeyer:2019ovu}. 
The question arises about
an additional, early process which can produce heavy $r$-process nuclei.
An interesting phenomenon could be the NS disruption by PBHs, as pointed out in Ref. \cite{Fuller:2017uyd,2019ASSP...56...91K}, which provides a viable site for $r$-process nucleosynthesis. For a discussion of the primordial black hole-neutron star destruction scenario as a possible "exotic" scenario for the origin of the heaviest elements see also \cite{2021RvMP...93a5002C}. New ideas to explain the origin of the $r$-process elements have also been evoked in recent works by Kirby et al. \cite{2023ApJ...958...45K} or Puls et al. \cite{2025A&A...693A.294A}.
The latter wrote: "Our results are challenging to explain from a nucleosynthetic point of view: the observationally derived abundances indicate the need for an additional early, primary formation channel (or a non-robust $r$-process)."

In Ref. \cite{2024arXiv241100535R}, also the possibility of IBBN as the
site where these conditions can occur was discussed.
In our work, primordial high-density clumps bound by gravitation are
considered, for instance, as an envelope of PBH.
These structures are sites where in the early Universe heavy nuclei can be produced if these structures are present after $t_{\rm nucl}$.
In contrast to other scenarios that consider unconventional objects such as magnetar giant flares \cite{Patel:2025frn}, magnetorotationally driven supernovae \cite{Winteler:2012hu}, or quark deconfinement supernovae
\cite{Fischer:2020xjl} as very early sources to produce the first amount of heavy nuclei, hot and dense matter is present from the beginning and must not be created by accretion from low-density, metal-free pop III matter as in the HBBN.

While details of the cosmic evolution of the Lagrange parameters
(temperature, density and proton fraction) in the context of IBB
nucleosynthesis scenarios still need to be worked out, there are recent
investigations of non-standard expansion scenarios as a basis for early
matter dominated phases with secondary (or multiple) inflation epoch(s) in
the early Universe. 
For details, see the recent review of possible expansion
histories by Allahverdi {\it et al.} \cite{2021OJAp....4E...1A}.
One particular example for an alternative to the standard homogeneous
Big Bang nucleosynthesis scenario to explain the observed
baryon-to-photon ratio $n_B/s\simeq 0.9\times 10^{-10}$ is to generate
an initially large baryon asymmetry
$\mu/T\sim 1$ -- $100$ from Affleck--Dine baryogenesis
\cite{Affleck:1984fy} that is diluted to the value currently observed
after the QCD phase transition \cite{Boeckel:2011yj}.\\

\section{Summary and Conclusions}
\label{sec:summary}
The QGP phase of the very early Universe is subject to variations in its EoS. Starting at extremely high temperatures, as it expands and cools down over time, particle species differentiate themselves from one another when the Universe's mean temperature falls below their respective masses. Eventually, the EoS value dips every time the "cosmic soup" reaches temperatures on the scale of the three SM interactions. We focused on studying the QCD transition from 1 MeV to 1300 MeV, when the Universe undergoes a phase change from QGP to hadron resonance gas. Unlike previous studies, we used a microscopical model of strongly interacting matter rather than lattice QCD. We compared several cosmic scenarios including different numbers of quark flavors, finding that the $b$ quark influences the cosmic EoS in temperature regimes as low as $T<1300$ MeV.

In light of several hints for physics Beyond the Standard Model, we included the putative X17 scalar boson in the calculation and compared the SM and SM+X17 scenarios. Our plots revealed a not negligible effect at low temperatures and some slight mitigation of QCD-induced dips. Although neither the X17 anomaly nor other SM anomalies have been firmly established as signs of new physics, the relevance of this study lies in exploring their potential cosmological implications. If such anomalies were to produce unrealistic cosmological effects, studies like ours could serve as a way to constrain or even rule out their existence.

For the past seven years, PBH distributions induced by the cosmic EoS have revealed an exciting number of features naturally emerging from cosmic history, unveiling peaks in the distribution at each of the cosmic phase transitions. We transformed our cosmic EoS to a PBH distribution normalized to account for 10\% of DM. Comparison of SM and SM + X17 scenarios revealed a long-lasting influence of X17. 
The secondary shoulder in the intermediate mass range induced by the X17 boson was expected from the behavior of the cosmic EoS. 
The PBH distribution revealed a long-lasting influence of the X17 at high horizon masses resulting from its influence in the low-temperature region of our study. 

Our model contains several important limitations that merit acknowledgment. First, we have assumed negligible lepton and baryon asymmetries throughout our analysis. This represents a significant simplification, as the Universe exhibits a baryon asymmetry of $\mathcal{O}(10^{-9})$, with lepton asymmetry expected to be of comparable magnitude. While standard scenarios predict such values, more exotic models propose substantially higher lepton asymmetries concentrated in the neutrino sector.
The work of Bödeker \cite{Bodeker:2020stj} demonstrates that these asymmetries can significantly modify both the cosmic equation of state and the resulting PBH mass distributions. Incorporating realistic asymmetry values into our framework represents a crucial avenue for future research that could substantially refine our predictions.

Additionally, our transformation from the cosmic equation of state to PBH mass distributions relies on simplified analytical prescriptions that introduce systematic uncertainties. Recent advances in computational methods offer promising alternatives. State-of-the-art relativistic simulations of gravitational collapse by Escrivà et al. \cite{Escriva:2019nsa,Escriva:2021aeh,Escriva:2022bwe} provide more accurate PBH formation probabilities and mass distributions. Integrating such sophisticated numerical approaches with our thermodynamic modeling would enhance the reliability of our theoretical framework and enable more robust comparisons with observational constraints.

We summarize several relevant constraints and circumstantial pieces of evidence regarding PBHs, focusing in particular on the intermediate-mass range, as our SM+X17 scenario predicts a shoulder in this region. This mass range appears to be relatively unconstrained, or at least the existing constraints are still subject to debate. Intermediate-mass PBHs could have significant implications for various cosmological and astrophysical phenomena.
Future observations from gravitational wave detectors, dynamical measurements from Gaia, or studies of dwarf galaxies may help to better constrain this mass window. Refinements of the theoretical BH accretion model could also have major implications on PBH constraints. We emphasize that our scenario could be either ruled out or confirmed within the next decade.
We also acknowledge that we have presented an optimistic PBH-DM interpretation of several astrophysical observations. PBHs are certainly not the only viable explanation for these phenomena. Similarly, we have interpreted many existing constraints as weak or non-threatening to our model. We consider the ongoing debate surrounding several of these constraints as a justification for keeping our approach open.

The existence of strong fluctuations in the gravitational field, for instance in connection with PBH, has strong consequences for the evolution of the early Universe. As an example, large hot and dense clumps of matter bound by these fluctuations appear as a new site where $r$-processes can occur so that it opens new possibilities to explain the early production of heavy elements. 

\subsection*{Acknowledgments}
The authors thank Dr. Florian Kühnel, Dr. Alberto Escriva, and Prof. Maxim Khlopov for their interest in our work and valuable discussions.
D.B. and O.I. were supported by the Polish NCN
under grant No. 2021/43/P/ST2/03319. G.R. acknowledges a joint stipend from the Alexander von Humboldt Foundation and the Foundation for Polish Science.

\appendix
\section{QCD transition and QGP (2+1) equation of state}
\label{app:QGP}
In this appendix we summarize the essence of the unified model \cite{Blaschke:2023pqd} for the QCD transition from a hadron resonance gas (HRG) to a QGP, where the hadrons are described as quark bound states (multiquark clusters) that can undergo a Mott dissociation, 
\begin{equation} 
\label{eq:omega_total_appendix}
    \Omega(T,\mu,\phi,\bar{\phi})=\Omega_{\rm QGP}(T,\mu,\phi,\bar{\phi})+\Omega_{\rm MHRG}(T,\mu,\phi,\bar{\phi}).
\end{equation}
The QGP part is described by a Polyakov loop Nambu--Jona--Lasinio (PNJL) model for the non-perturbative low-energy domain augmented by perturbative QCD corrections $\mathcal{O}(\alpha_s)$,
\begin{equation}
\Omega_{\rm QGP}(T,\mu,\phi,\bar{\phi})=\Omega_{\rm PNJL}(T,\mu,\phi,\bar{\phi})+\Omega_{\rm pert}(T,\mu,\phi,\bar{\phi}).
\end{equation}
The MHRG part takes the form of a cluster decomposition of the thermodynamic potential for quark matter. 
\begin{eqnarray}
\label{eq:Omega}
\Omega_{\rm MHRG}(T,\mu,\phi,\bar{\phi}) &=& \sum_{n=2}^{N}  \Omega_n(T,\mu) + \Phi\left[\left\{S_n \right\} \right],
\\
\Omega_n(T,\mu) &=& c_n\left[{\rm Tr} \ln S_n^{-1} + {\rm Tr} (\Pi_n S_n) \right],
\label{eq:Omega-a}
\end{eqnarray}    
where $n$ denotes the total number of valence quarks and antiquarks in the cluster; $c_n=1/2$ for bosonic and $c_n=-1$ for fermionic clusters \cite{Vanderheyden:1998ph,Blaizot:2000fc}. 
The functional $\Phi\left[\left\{S_n \right\} \right]$ contains all two-cluster irreducible (2CI) closed-loop diagrams that can be formed with the complete set of Green's functions of the cluster $S_n$. 
We will restrict ourselves to a maximum number of $N=6$ quarks in the cluster and to the class of two-loop diagrams of the "sunset" type; see Ref. \cite{Blaschke:2023pqd} for details.

The density is obtained in the form of a generalized
Beth-Uhlenbeck EoS
\begin{eqnarray}
\label{eq:n}
n_{\rm MHRG}(T,\mu)&=& \sum_i a_i \, d_i\, c_{a_i}\int \frac{d^3q}{(2\pi)^3} 
\int_0^\infty \frac{d\omega}{\pi}
\left\{f^{(a_i),+}_\phi \right.
\nonumber\\
&&\left.-\left[f^{(a_i),-}_\phi\right]^*\right\}
2 \sin ^2 \delta_{n_i}(\omega,q) \frac{\partial  \delta_{n_i}(\omega,q)}{\partial \omega} ~,
\nonumber\\
\end{eqnarray}
where the properties of the distribution function $f^{(a),+}_\phi$ and the phase shift with respect to 
reflection $\omega \to -\omega$ have been used and the 
"no sea" approximation has been employed which removes the divergent vacuum contribution.
The Polyakov--loop modified distribution functions are defined as
\begin{eqnarray}
\label{eq:PL-function}
    f^{(a),\pm}_{\phi}~\stackrel{\text{(a even)}}{=}&&\frac{({\phi} - 2\bar{\phi} y_a^\pm) y^\pm_a + {y_a^\pm}^3}{1 - 3 ({\phi} - \bar{\phi} y_a^\pm) y_a^\pm - {y_a^\pm}^3}~,\\
    f^{(a),\pm}_{\phi}~\stackrel{\text{(a odd)}}{=}&&\frac{(\bar{\phi} + 2\phi y_a^\pm) y^\pm_a + {y_a^\pm}^3}{1 + 3 (\bar{\phi} + \phi y_a^\pm) y_a^\pm + {y_a^\pm}^3}~,
\end{eqnarray}
where $y^\pm_a=e^{-\left( \omega \mp a\mu\right)/T}$ 
and $a$ is the net number of valence quarks present in the cluster. 
See Appendix A of Ref. \cite{Blaschke:2023pqd} for a detailed derivation.

In an analogous manner follows for the MHRG entropy density
\begin{eqnarray}
s_{\rm MHRG}(T,\mu)&=& -\frac{\partial \Omega}{\partial T} 
= \sum_i s_i(T,\mu) \nonumber\\
&=& \sum_i  d_i \, c_{a_i}\int \frac{d^3q}{(2\pi)^3}\int \frac{d\omega}{\pi}
\left\{
\sigma^{(a_i),+}_\phi \right.
\nonumber\\
&&\left.
+\left[\sigma^{(a_i),-}_\phi\right]^*
\right\}
 2 \sin ^2 \delta_{n_i}(\omega,q) \frac{\partial  \delta_{n_i}(\omega,q)}{\partial \omega} 
~,
\nonumber\\
\label{eq:s}
\end{eqnarray} 
where $\sigma^{(a)} =  f^{(a)}_\phi  \ln f^{(a)}_\phi (-)^a [1(-)^a f^{(a)}_\phi] \ln [1(-)^a f^{(a)}_\phi]$ and $f^{(a)}_\phi$ is the cluster distribution function for a net quark number $a$ modified by the traced Polyakov loop.
The formula for the pressure as thermodynamical potential can be obtained from Eq. (\ref{eq:n}) by integration  over the quark chemical potential $\mu$.
Analogously, it can be obtained from Eq. (\ref{eq:s}) by integration over $T$
\begin{equation}
    \label{eq:pressure}
    p(T,\mu) = \int_0^T dT' s(T',\mu)~.
\end{equation}

\section{Details on the Cosmic EoS Computation}
\label{app:CosmicEoS}
This section summarizes the method presented in the annex of
\cite{Borsanyi:2016ksw},
accompanying ref. \cite{Borsanyi:2016ksw},
to obtain a cosmic EoS from the lattice QCD result.
We used 147 points calculated in the microscopical model \cite{Blaschke:2023pqd} with $T\in [1;1300]$ MeV with a temperature step $dT=1$ MeV for $T\in [1;19]$ MeV and $dT=10$ MeV for $T\in [30;1300]$ MeV to retrieve the QGP (2+1) pressure, entropy and energy density $p_{\rm QGP}^{(2+1)}$, $s_{\rm QGP}^{(2+1)}$, and $\varepsilon_{\rm QGP}^{(2+1)}$, respectively. 
Since the temperature step changes, we want to avoid using numerical derivation/integration on thermodynamic data.

The simplest species contribution to introduce are the photons and neutrinos: $p_\gamma = \pi^2T^4/45$ for 3 lepton flavors considering their antiparticle $p_\nu=(7/8)\times2\times3\times \pi^2T^4/90$.

Charged lepton and quark contributions both use the same formula of free energy density of fermion fields with masses $m_i$ and the spin degeneracy factor $g_i$: 
\begin{equation}\label{eq:fermionContribution}
    \frac{p(T, m_i)}{T^4} = -\frac{1}{2\pi^2}\sum_{i}g_i \left(\frac{m_i}{T}\right)^2 \sum_{k=1}^{\infty} \frac{(-1)^k}{k^2}K_2 \left(\frac{ k \,m_i}{T}\right)
\end{equation}
with $K_2$ the modified Bessel function of the 2nd kind, $g_i$ the spin degeneracy factor $g_l=4$ for leptons, and $g_q=12$ for quarks. $i=l \in (e, \mu, \tau)$ give the pressure contribution of the charged leptons. 
Using equation (\ref{eq:PerfectFluid}) an expression of entropy density can be derived as a function of temperature and field mass $m_i$. 

The bosonic contribution is the following. 
\begin{equation}\label{eq:bosonContribution}
    \frac{p}{T^4} = \frac{1}{2\pi^2}\sum_{i}g_i \left(\frac{m_i}{T}\right)^2 \sum_{k=1}^{\infty} \frac{1}{k^2}K_2 \left(\frac{k\,m_i}{T}\right),
\end{equation}
where $g_i=3$ for $W^\pm$ and $Z^0$ and $g_i=1$ for $H^0$.

The charm correction for the third quark mass is: 
\begin{equation}\label{eq:charmCorrection}
    \frac{p^{(2+1+1)}(T)}{P^{(2+1)}(T)}= \frac{\sigma_{\rm SB}(3)+p_{\rm charm}(T)}{\sigma_{\rm SB}(3)},
\end{equation}
where $\sigma_{\rm SB}(n_f)$ the Stefan-Boltzmann limit for $n_f$ quark flavor considered and $p_{\rm charm}(T)$ from (\ref{eq:fermionContribution}) with $m_i=m_c$. 

The bottom correction for the fourth quark mass is similar:
\begin{equation}\label{eq:bottomCorrection}
    \frac{p^{(2+1+1+1)}(T)}{p^{(2+1+1)}(T)}= \frac{\sigma_{\rm SB}(4)+p_{\rm bottom}(T)}{\sigma_{\rm SB}(4)}
\end{equation}
$p_{\rm bottom}(T)$ from (\ref{eq:fermionContribution}) with $m_i=m_c$. 
Note that the above equations (\ref{eq:charmCorrection}) and (\ref{eq:bottomCorrection}) apply on $p_{\rm QGP}$ and not on $p_{cosmic}$.

\section{Details on the horizon mass calculation}\label{app:horizonMass}

In this section, we discuss the calculation of the horizon mass based on equation \eqref{eq:horizonMass}. As it depends on $g_\varepsilon$ (see Equation \eqref{eq:degreesOfFreedom}), the horizon masses for the SM and SM+X17 scenarios exhibit slight differences.

\begin{figure}[t]
    \centering
    \includegraphics[width=\linewidth]{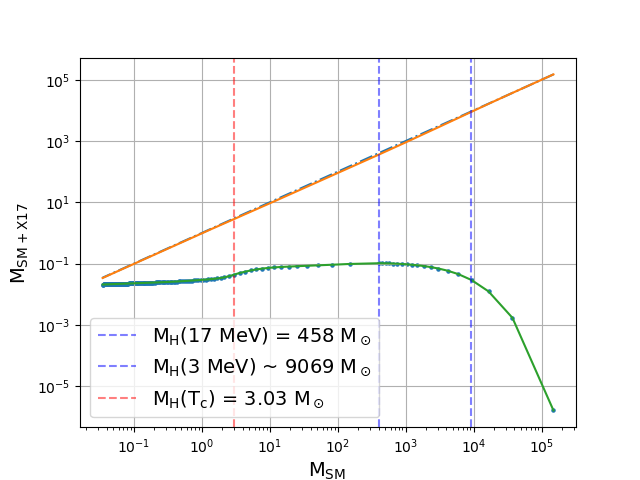} 
    \caption{Comparison of the horizon mass for the 2 models, equation \ref{eq:horizonMass}. The orange curve shows the horizon mass of SM+X17 model as a function of that for the SM model. The green curve displays the relative difference between the 2 models and the dash dotted line stands for  
    $M_{\rm SM+X17}=M_{\rm X17}$ to guide the eyes. 
    }
    \label{fig:horizonMassComparaison}
\end{figure}

\begin{figure}[t]
    \centering
    \includegraphics[width=\linewidth]{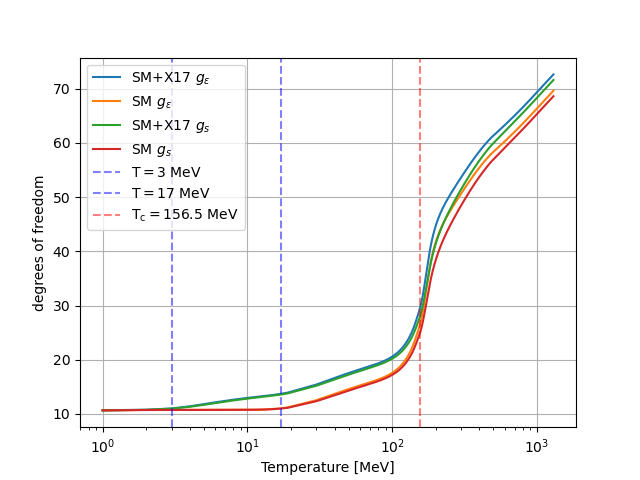} 
    \caption{Relativistic degrees of freedom for a (2+1+1+1) quark flavors configuration for the SM and SM+X17 model, see equation \eqref{eq:degreesOfFreedom}.}
    \label{fig:degreesOfFreedomComparison}
\end{figure}

The relative difference between the two cases is given by:
\begin{equation}
\label{eq:horizonMassRelativeDifference}
\frac{M_{\rm SM}-M_{\rm X17}}{M_{\rm SM}} = 1 - \sqrt{\frac{g_{\rm SM}}{g_{\rm SM+X17}}}
\end{equation}
where $g_{\rm SM+X17}=g_{\rm QGP}+g_\gamma+g_\nu +g_{e, \mu, \tau}+g_{W^\pm, Z^0, H^0}+g_{\rm X17} = g_{\rm SM}+g_{\rm X17}$. The values of $g_{\rm SM}$ and $g_{\rm SM+X17}$ are shown in figure \ref{fig:degreesOfFreedomComparison}. As illustrated in Figure \ref{fig:horizonMassComparaison}, the difference in horizon mass between the two models remains small, the orange curve closely following the diagonal line at all temperatures. The relative difference remains below $\rm 10\%$ throughout.

Using equation \eqref{eq:horizonMassRelativeDifference} along with figures \ref{fig:horizonMassComparaison} and \ref{fig:degreesOfFreedomComparison}, we identify four thermodynamic regimes that characterize the behavior of the universe:

\begin{itemize}
    \item \textbf{Low temperature / high horizon mass} $\rm T\lesssim 3\,MeV$: Below the temperature for the onset of the X17 particle at $\rm T\lesssim 17\,MeV/6 \sim 3\,MeV$, the cosmic EoS consists of only massless particles and the effective number of degrees of freedom is constant $g_{\rm SM} \sim g_{\rm SM+X17}$. 
    The relative difference in the horizon mass falls to very low values, and PBHs formed at this stage have nearly identical masses in both models.
    
    \item \textbf{X17 appearance} $\rm 17\,MeV/6 < T <17\,MeV$: This is the region where the X17 particle begins to contribute to the cosmic EoS (as discussed in Section \ref{ssec:CosmicEoS}), and the relative difference in the horizon masses of both models reaches $10\%$. PBHs in the SM scenario become slightly more massive than in the SM+X17.

    \item \textbf{Massive SM particle appearance} $\rm 17\,MeV<T<T_c$: Here, in addition to the massless particles (radiation) and the X17 particle, the lightest SM particles appear according to the rule for their threshold temperatures are muons with $T_\mu \sim m_\mu/6 = 17.6$ MeV and pions with  $T_\pi \sim m_\pi/6 = 23.3$ MeV. Accounting for these lightest massive particles approximately doubles the effective number of degrees of freedom and thus the pressure (see Figure \ref{fig:PartialPressure}), with $g_{\rm SM+X17} > g_{\rm SM}$. The relative difference reaches a plateau around $\rm 10\%$. In this temperature range, the PBHs formed in the SM scenario are $\sim10\%$ more massive than those formed in the SM+X17 model.

     \item \textbf{QGP appearance / low horizon mass} for $ T \gtrsim T_c$: In this regime, the QGP thermodynamics dominates (see Figure \ref{fig:PartialPressure}), leading to $g_{\rm SM} \sim g_{\rm SM+X17}$. As a result, the relative difference in horizon mass between the two models is minimal, and PBHs formed at this stage have similar masses in both scenarios.
    
\end{itemize}

\section{More PBH mass distributions}\label{app:PBHdistribution}
\begin{figure}[t]
    \centering
    \includegraphics[width=\linewidth]{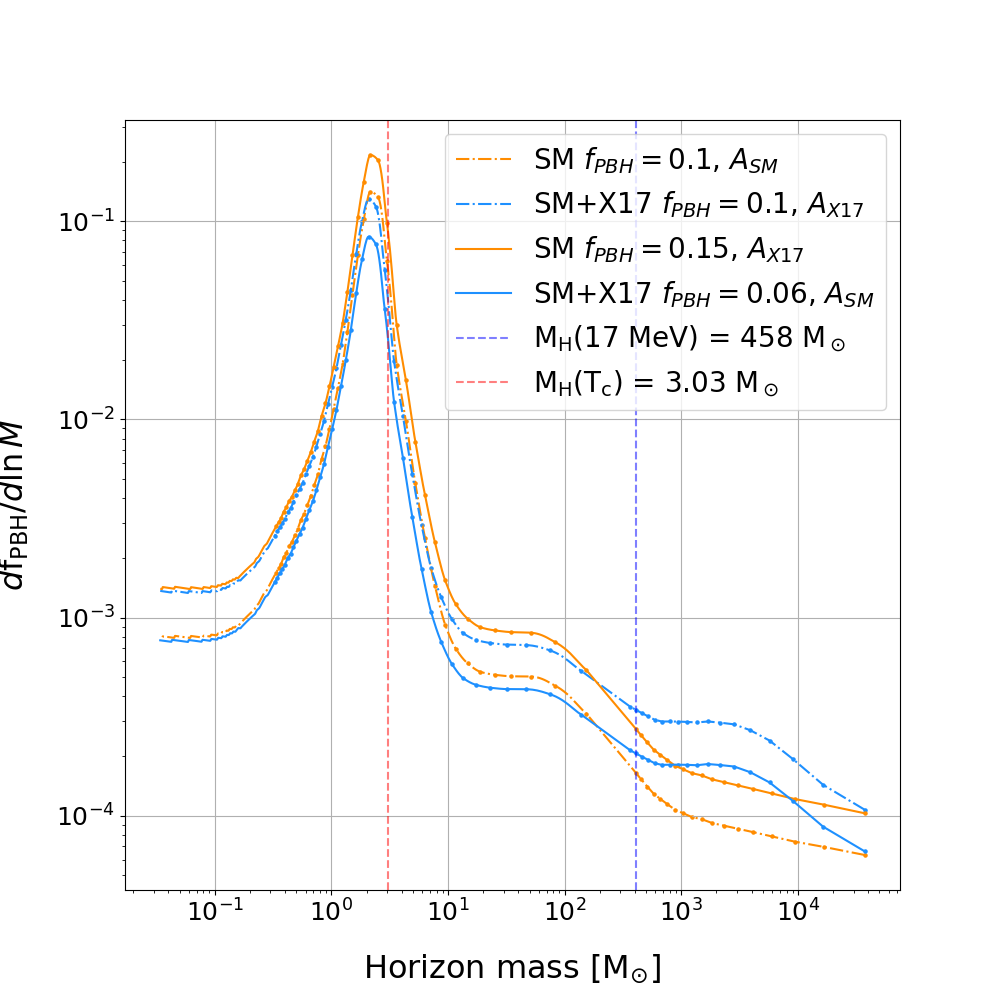} 
    \caption{PBH distribution using different fluctuation amplitudes $\rm A_{opt}^{SM}=0.0614$ and $\rm A_{opt}^{X17}=0.0621$. Dash-dotted lines are PBH distributions normalized at $f^{i}_{\rm PBH}({\rm A}^{i})=0.1$ where $i=$ SM, SM+X17; continuous lines are normalized with $f^{i}_{\rm PBH}({\rm A}^{j})=0.1$, where $i\neq j$.}
    \label{fig:all_pbh_distribution}
\end{figure}

In figure \ref{fig:all_pbh_distribution}, we display the PBH distribution as a function of the horizon mass for different normalization points $\rm A_{\rm opt}$. The blue dashed-dotted line shows $f^{\rm X17}_{\rm PBH}({\rm A^{X17}})=0.1$, while the continuous orange line shows $f^{\rm SM}_{\rm PBH}({\rm A^{X17}})=0.15$. The orange dashed-dotted line for $f^{\rm SM}_{\rm PBH}({\rm A^{SM}})=0.1$ can be compared with the blue continuous line for $f^{\rm X17}_{\rm PBH}({\rm A^{SM}})=0.06$, and the characteristics of the figures \ref{fig:CosmicEoS} and \ref{fig:criticalDensity} are retrieved.  


\end{document}